\newif\ifblind
\blindfalse

\newif\iflong
\longtrue
\documentclass{llncs}

\usepackage{amsmath}
\usepackage{graphicx}
\usepackage{subcaption}
\usepackage{booktabs}
\usepackage[table,dvipsnames]{xcolor}
\usepackage{url}
\usepackage{array}

\usepackage{breakurl}
\usepackage[breaklinks]{hyperref}
\hypersetup{pdfborderstyle={/S/U/W 1},colorlinks,urlcolor=blue,citecolor=blue}

\graphicspath{{figs/}}
\DeclareGraphicsExtensions{.png,.pdf}

\newcommand\mytt[1]{\texttt{\small{#1}}}

\begin{document}

\date{}

\title{\Large \bf A Comparison of SynDiffix Multi-table versus Single-table Synthetic Data}
\titlerunning{Query-based Analysis}

\ifblind
\author{
First Last, First Last, First Last \\
Affiliations \\
emails
}
\else
\author{Paul Francis}
\institute{Max Planck Institute for Software Systems (MPI-SWS)}
\fi

\maketitle

% Use the following at camera-ready time to suppress page numbers.
% Comment it out when you first submit the paper for review.
%\thispagestyle{empty}

%\subsection*{Abstract}

\begin{abstract}
 
SynDiffix is a new open-source tool for structured data synthesis. It has anonymization features that allow it to generate multiple synthetic tables while maintaining strong anonymity. Compared to the more common single-table approach, multi-table leads to more accurate data, since only the features of interest for a given analysis need be synthesized. This paper compares SynDiffix with 15 other commercial and academic synthetic data techniques using the SDNIST analysis framework, modified by us to accommodate multi-table synthetic data. The results show that SynDiffix is many times more accurate than other approaches for low-dimension tables, but somewhat worse than the best single-table techniques for high-dimension tables.

\end{abstract}  

\section{Introduction}
\label{sec:intro}

In recent years there has been a lot of interest in synthetic structured data for statistical disclosure control. The US National Institute of Standards and Technology (NIST) has a research program~\cite{nistCrc} to better understand the utility and privacy of data deidentification mechanisms. They have developed the SDNIST software tool and several sample datasets to analyze and compare the utility and privacy of synthetic data mechanisms\footnote{https://github.com/usnistgov/sdnist}. NIST has so far archived the test results of nearly 30 different techniques from more than a dozen organizations.

There are a variety of use cases for synthetic data, including for instance data enhancement and generating test data. These use cases do not necessarily require that the synthetic data very accurately mimics the original data. Indeed sometimes the goal is to modify the statistics of the original data, for instance to remove bias or increase certain profiles.

The primary use case for SDNIST, however, is statistical disclosure. The goal is to replicate the original data as accurately as possible while preserving anonymity. The original datasets supplied by SDNIST come from US Census American Community Survey (ACS) data, and the SDNIST utility metrics measure how closely the synthetic data matches the original data.

Most synthetic data techniques are designed to produce \emph {single-table} datasets---all columns of the original dataset are synthesized to produce a single synthesized dataset. An advantage of single-table approaches is ease of use. The single table can be pulled into any data analysis tool and manipulated exactly as with the original data. An important disadvantage of single-table approaches, however, is that data accuracy degrades with increased columns.

An alternate approach is to make \emph{multi-table} datasets. Each table synthesizes only the columns necessary for a given analytic task. For instance, suppose there are 50 columns in the data, and the analyst is interested in the correlation between two columns. In the single-table approach, nominally all 50 columns would be synthesized, and the two columns of interest taken from the 50-column synthesized table. In a multi-table approach, a synthesized table with only those two columns would be built.

In the multi-table approach thousands of tables, one for each analytic question, can easily be required. Multi-table is therefore only viable if it is strongly anonymous despite multiple tables. Prior synthetic data mechanisms do not have this characteristic. The more tables that are generated from the same data, the more privacy is lost. It is unknown how many tables can be built before anonymity is dangerously degraded, because existing systems are not designed to be used in multi-table mode, and therefore such testing is unnecessary.

%zzzz TODO check recent attack paper zzzz

Although multi-table is certainly less convenient than single-table, it is common practice among statistics offices to release data as multiple tables, each with a relatively small number of columns.

\begin{figure}
\begin{center}
\includegraphics[width=1.0\linewidth]{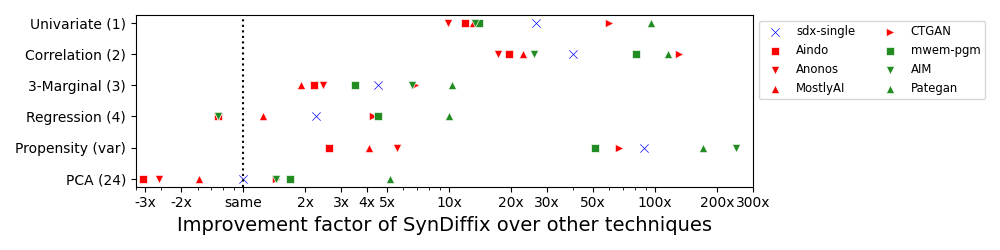}
\caption{Improvement factor of SynDiffix over other techniques for each measure (number of measured columns). Techniques with insufficient anonymity or fewer than 24 columns in synthetic table are not comparable and are therefore excluded. Measures with a negative improvement factor (left of the dashed line) are better than SynDiffix. Measures greater than 300x are not shown. Note log scale.}
\label{fig:summaryPlotLog}
\end{center}
\end{figure}

SynDiffix~\cite{francis2023syndiffix} is a new structured synthetic data generator that remains strongly anonymous no matter how many tables are generated (\S\ref{sec:syndiffix}). This paper uses SDNIST to compare the utility and privacy of SynDiffix against 15 other mechanisms, each of which have been submitted to SDNIST by its respective developers. We show that SynDiffix is many times more accurate than the best alternative techniques for low-dimensional tables (see Figure~\ref{fig:summaryPlotLog}).  Compared to the next best techniques, which are proprietary commercial products, SynDiffix' median measure is many times more accurate for low-dimension measures: 10x more accurate for single-column measures, 17x more accurate for 2-column measures, and 2x more accurate for 3-column measures.

That advantage degrades for higher-dimension measures. For a 4-column measure (linear regression), SynDiffix is more accurate than most techniques, but 30\% less accurate than the best techniques. For the 24-column measure (PCA), SynDiffix is again more accurate than most techniques, but 3x less accurate than the best technique.

We also show that, using SDNIST's simple privacy metric, SynDiffix has very strong anonymity. While most of the mechanisms also have strong anonymity, SynDiffix has even stronger anonymity than generative approaches, and slightly weaker anonymity than the differential privacy mechanisms.

The contributions of this paper are:
\begin{itemize} 
    \item A broad comparison of utility for SynDiffix' multi-table synthesis with 15 single-table synthetic data techniques using SDNIST.
    \item An improved interpretation of SDNIST's privacy metric, taking into account the statistical baseline of the data.
    \item An enhancement of the SDNIST tool to accommodate multi-table approaches.
\end{itemize} 

\textbf{Limitations:} While the measures in this paper do allow for a direct comparison of techniques, they do not measure utility for actual use cases. SynDiffix' superior (or inferior) accuracy may or may not be important for any given use case. This paper only measures one dataset. Measures with other datasets, for a subset of the techniques in this paper, can be found in~\cite{francis2023syndiffix}, and reinforce the findings of this paper.

\section{Overview of SynDiffix}
\label{sec:syndiffix}

Here we give a brief overview of SynDiffix. A full description can be found at~\cite{francis2023syndiffix}.

SynDiffix operates by building multi-dimensional search trees from the original data, and then assigning synthetic data from the nodes of the search trees. SynDiffix builds a family of trees with different dimensions: all one-dimension trees, all two-dimension trees, up to a single tree with all columns. Lower-dimension trees have better precision, while higher-dimension trees better capture the relationship between attributes. SynDiffix uses information from all trees to synthesize data. If the original data has too many columns to build all combinations of all dimensions, SynDiffix partitions the original table into multiple lower-dimension tables, synthesizes each table, and then merges them back together.

SynDiffix' anonymity derives from how it assigns nodes in the trees. The nodes themselves incorporate three core anonymization features, \textit{range snapping, sticky noise, and aggregation}.

\textbf{Range snapping:} Unlike most search trees, which partition ranges by splitting them in half or in a way that balances the tree, SynDiffix forces ranges to conform to a fixed set of sizes and offsets. Specifically, ranges conform to a power-of-two sequence (\mytt{..., 1/4, 1/2, 1, 2, 4, ...}), and likewise for offsets (a range of 2 can fall on offsets \mytt{..., -2, 0, 2, 4, ...}). Text and datetime values are converted to numbers prior to tree building, and converted back when values are assigned from the tree nodes.

\textbf{Aggregation:} All nodes in all trees are aggregates of multiple individuals. Nodes without enough individuals are pruned from the trees. For time-series data, SynDiffix is configured to identify which rows belong to which individuals.

\textbf{Sticky noise:} Noise is added to each node's row count. The noise is sticky in that for any given node, as defined by the ranges and offsets of each attribute, the amount of noise is always the same.

Taken together, these three mechanisms lead to strong anonymization. Aggregation ensures that no individual data is released. Range snapping and sticky noise ensures that any given data value will always appear in the same limited set of aggregates with the same noise, no matter how many synthetic tables are generated. In this sense, SynDiffix is similar to the Cell Key Method used by the statistics offices of Australia~\cite{thompson2013methodology}, Germany~\cite{geyer2022perspectives}, and the UK~\cite{onsCellKey}. SynDiffix, however, generalizes the approach to work automatically, including with time-series data, and to produce synthetic data rather than tabular data.

\section{Setup}
\label{sec:background}

The SDNIST synthetic data measurement tool works with three datasets. Each is composed of 24 attributes from the 44-question ACS~\cite{acsSampleForm2020}. For this study, we use the \mytt{NATIONAL} dataset, which consists of 27253 rows over 20 PUMA regions~\cite{acsGeoConcepts} across the USA.

The 16 techniques measured in this paper are listed in Table~\ref{tab:infotable}. Five of the techniques use differential privacy (DP), colored green. Several of these use generative modeling, and five other techniques use generative modeling without DP, colored red. Five techniques do not synthesize all 24 columns, and we regard four techniques as having excessively weak anonymity (\S\ref{sec:privacy}). As these are not directly comparable, they have lighter color shades so that they can be distinguished in the various plots.

\begin{table}
    \centering
    \caption{Set of compared techniques, showing the number of columns synthesized (out of 24), and whether or not anonymization is weak. Technique labels link to the SDNIST report. Techniques without an epsilon do not use differential privacy. Techniques without both a repo and citation are proprietary.}
    \label{tab:infotable}
    \begin{tabular}{lllllllll}
        \toprule
          & Technique & Tech & Org & Cols & Weak & \thinspace$\epsilon$\qquad\qquad & Cite & Repo \\
        \midrule
        \cellcolor{blue} & \href{https://htmlpreview.github.io/?https://github.com/yoid2000/sdnist-summary/blob/main/results/syndiffix_all/report.html}{SynDiffix} & K-dimension search trees & \href{https://www.open-diffix.org/}{Open Diffix} & 24 &   &  & \cite{francis2023syndiffix} & \href{https://github.com/diffix/syndiffix}{link} \\
        \cellcolor{blue} & \href{https://htmlpreview.github.io/?https://github.com/yoid2000/sdnist-summary/blob/main/results/sdx-single/report.html}{sdx-single} & K-dimension search trees & \href{https://www.open-diffix.org/}{Open Diffix} & 24 &   &  & \cite{francis2023syndiffix} & \href{https://github.com/diffix/syndiffix}{link} \\
        \cellcolor{red} & \href{https://htmlpreview.github.io/?https://github.com/yoid2000/sdnist-summary/blob/main/results/aindo_synth/report.html}{Aindo} & Generative model & \href{https://www.aindo.com/}{Aindo} & 24 &   &  &  &  \\
        \cellcolor{red} & \href{https://htmlpreview.github.io/?https://github.com/yoid2000/sdnist-summary/blob/main/results/anonos_sdk/report.html}{Anonos} & Generative model & \href{https://www.anonos.com/}{Anonos} & 24 &   &  &  &  \\
        \cellcolor{red} & \href{https://htmlpreview.github.io/?https://github.com/yoid2000/sdnist-summary/blob/main/results/mostlyai_sd_platform/report.html}{MostlyAI} & Generative model & \href{https://mostly.ai/}{MostlyAI} & 24 &   &  &  &  \\
        \cellcolor{red} & \href{https://htmlpreview.github.io/?https://github.com/yoid2000/sdnist-summary/blob/main/results/sdv_ctgan_epochs1000/report.html}{CTGAN} & Generative model & \href{https://sdv.dev/}{SDV} & 24 &   &  & \cite{xu2019modeling} & \href{https://github.com/sdv-dev/SDV}{link} \\
        \cellcolor{pink} & \href{https://htmlpreview.github.io/?https://github.com/yoid2000/sdnist-summary/blob/main/results/ydata_fabric_synthesizers/report.html}{YData} & Generative model & \href{https://ydata.ai/}{YData} & 24 & X &  &  & \href{https://github.com/ydataai/ydata-synthetic}{link} \\
        \cellcolor{ForestGreen} & \href{https://htmlpreview.github.io/?https://github.com/yoid2000/sdnist-summary/blob/main/results/mwem_pgm/report.html}{mwem-pgm} & Graphical models + DP & \href{https://dream.cs.umass.edu/}{See pub} & 24 &   & 1 & \cite{mckenna2019graphical} &  \\
        \cellcolor{ForestGreen} & \href{https://htmlpreview.github.io/?https://github.com/yoid2000/sdnist-summary/blob/main/results/aim_e_10_all/report.html}{AIM} & Workload adaptive + DP & \href{https://opendp.org/}{OpenDP} & 24 &   & 10 & \cite{mckenna2022aim} & \href{https://github.com/opendp/smartnoise-sdk}{link} \\
        \cellcolor{ForestGreen} & \href{https://htmlpreview.github.io/?https://github.com/yoid2000/sdnist-summary/blob/main/results/pategan_n_iter_50_e_10_all/report.html}{Pategan} & Generative model + DP & \href{https://github.com/PerceptionLab-DurhamUniversity/pategan}{See pub} & 24 &   & 10 & \cite{jordon2018pate} & \href{https://github.com/PerceptionLab-DurhamUniversity/pategan}{link} \\
        \cellcolor{YellowGreen} & \href{https://htmlpreview.github.io/?https://github.com/yoid2000/sdnist-summary/blob/main/results/genetic_sd_e_10_simple/report.html}{Genetic} & Approximate DP & \href{https://github.com/giusevtr/private_gsd}{See pub} & 21 &   & 10 & \cite{liu2023generating} & \href{https://github.com/giusevtr/private_gsd}{link} \\
        \cellcolor{YellowGreen} & \href{https://htmlpreview.github.io/?https://github.com/yoid2000/sdnist-summary/blob/main/results/sarus_sdg_demographic/report.html}{Sarus} & Generative model + DP & \href{https://www.sarus.tech/}{Sarus} & 10 &   & 10 & \cite{canale2022generative} &  \\
        \cellcolor{SkyBlue} & \href{https://htmlpreview.github.io/?https://github.com/yoid2000/sdnist-summary/blob/main/results/cart_cf21/report.html}{CART} & Decision trees & \href{https://synthpop.org.uk/}{Synthpop} & 21 &   &  & \cite{nowok2016synthpop} & \href{https://CRAN.R-project.org/package=synthpop}{link} \\
        \cellcolor{Salmon} & \href{https://htmlpreview.github.io/?https://github.com/yoid2000/sdnist-summary/blob/main/results/k_anonymity_k_6/report.html}{K6-Anon} & K-anonymity & \href{https://github.com/sdcTools/sdcMicro}{sdcMicro} & 10 &   &  & \cite{templ2015statistical} & \href{https://github.com/sdcTools/sdcMicro}{link} \\
        \cellcolor{Goldenrod} & \href{https://htmlpreview.github.io/?https://github.com/yoid2000/sdnist-summary/blob/main/results/pram_default/report.html}{PRAM} & Random value changes & \href{https://github.com/sdcTools/sdcMicro}{sdcMicro} & 10 & X &  & \cite{meindl2019feedback} & \href{https://github.com/sdcTools/sdcMicro}{link} \\
        \cellcolor{Tan} & \href{https://htmlpreview.github.io/?https://github.com/yoid2000/sdnist-summary/blob/main/results/smote_target_marital/report.html}{SMOTE} & Minority oversampling & \href{https://github.com/ut-dallas-dspl-lab/AI-Fairness}{See pub} & 24 & X &  & \cite{zhou2023improving} & \href{https://github.com/ut-dallas-dspl-lab/AI-Fairness}{link} \\
        \cellcolor{Tan} & \href{https://htmlpreview.github.io/?https://github.com/yoid2000/sdnist-summary/blob/main/results/subsample_40pcnt_all/report.html}{Sample40} & Simple sampling & \href{https://www.census.gov/content/dam/Census/library/publications/2021/acs/acs_pums_handbook_2021_ch01.pdf}{US Census} & 24 & X &  & \cite{acsBasics2021} &  \\
        \bottomrule
    \end{tabular}
\end{table}

For SynDiffix, we generated 455 separate tables, one for each column combination measured by SDNIST. These include 24 1-column tables, 181 2-column, 232 3-column, one 4-column, 16 9-column, and one 24-column table. This amounts to 1254 total columns, making the storage requirements for SynDiffix 52 times that of a single table. We used the default settings for SynDiffix' average suppression threshold (5 rows), absolute suppression threshold (3 rows), and per-bucket noise standard deviation (1.4). It took just under 4 hours to generate all 455 synthetic tables on a windows laptop with 32G RAM and an Intel i7-7820HQ CPU running at 2.9GHz.

We additionally show the results for SynDiffix as a single table (`sdx-single'). This allows us to directly see the benefits of a multi-table approach, as well as allows us to compare SynDiffix with other techniques on an apples-to-apples usability basis.

\subsection{Changes to SDNIST}

The SDNIST tool assumes a single synthetic table. In order to measure SynDiffix, we modified SDNIST to handle multiple tables. The Github repo with these modifications is \href{https://github.com/yoid2000/SDNist-multi}{https://github.com/yoid2000/SDNist-multi}. Prior to each measurement operation, SDNIST-multi fetches the synthetic table that has the same columns required by the measurement. Other changes required by SDNIST-multi are discussed in the relevant sections.

\section{SDNIST measure results}
\label{sec:results}

This section presents the results of the SDNIST measures. The code used to produce these results is \href{https://github.com/yoid2000/sdnist-summary}{https://github.com/yoid2000/sdnist-summary}.

A summary of most of the utility measures is shown in Figure~\ref{fig:summaryPlotLog}, which give the \textit{improvement factor} IF of SynDiffix ($sdx$) over the other techniques ($alt$). IF is computed as:

\begin{equation}
    \text{IF} =
\begin{cases}
    \delta_{alt} / \delta_{sdx} & \text{if } \delta_{alt} \geq \delta_{sdx} \\
    -\delta_{sdx} / \delta_{alt}, & \text{otherwise}
\end{cases}
\label{eq:if}
\end{equation}

where $\delta_{x} = \lvert S_{perfect} - S_{actual} \rvert$, the absolute difference between a perfect score and the actual score for the technique $x$. The intuition here is that, compared to 10\% error, 5\% error represents an IF of 2x, and 2.5\% error has an IF of 4x. A negative value is used to denote the case where the other technique has a better score than SynDiffix.

The measures in Figure~\ref{fig:summaryPlotLog} are ordered by lowest to highest dimensionality, with univariate measuring only one column and PCA being a measure based on all 24 columns. Figure~\ref{fig:summaryPlotLog} excludes non-comparable techniques: those that either do not synthesize all 24 columns, or have weak anonymity. Those measures are however included in all other results.

Figure~\ref{fig:summaryPlotLog} shows that SynDiffix improves over other techniques by multiple factors for low-dimensional measures, but that the advantage degrades and is even reversed somewhat for high-dimensional measures.

\subsection{Privacy}
\label{sec:privacy}

To measure privacy, SDNIST emulates an attack whereby the attacker knows eight quasi-identifying attributes of an individual (\mytt{EDU, SEX, RAC1P, PUMA, OWN\_RENT, INDP\_CAT, HISP, MSP}), finds a record in the synthetic data that uniquely matches the quasi-identifier, and infers that the remaining attributes are those of the individual.

\begin{figure}
\begin{center}
\includegraphics[width=0.75\linewidth]{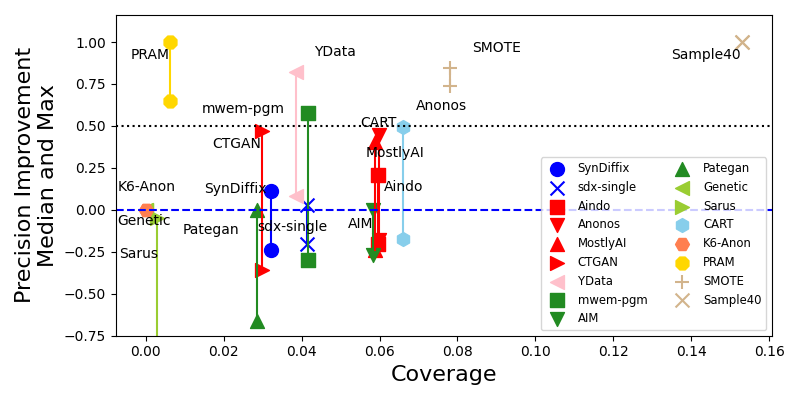}
\caption{Precision Improvement (PI) and Coverage where the attacker knows the quasi-identifiers of the target, finds a record with a unique and complete match of the quasi-identifiers, and infers an unknown attribute from that record. PI below 0.0 has no privacy loss whatsoever. PI below 0.5 has strong anonymity.}
\label{fig:attackPrecCovPairs}
\end{center}
\end{figure}

In a multi-table setting, this attack is most effective if the synthetic table contains the quasi-identifying columns plus only one additional column; the inferred column. Therefore, in measuring this attack for SynDiffix, we generated 16 separate 9-column tables, one for each inferred column.

Although SDNIST measures the precision of this attack for each column, it does not correctly interpret the privacy loss associated with the precision measure. It its summary reports (linked from Table~\ref{tab:infotable}), SDNIST claims that a 50\% average precision (``percentage of matched records'') represents strong privacy. The problem with this interpretation is that it does not take into account the statistical baseline of the data itself. 50\% inference precision represents substantial privacy loss for an attribute that occurs in say only 1\% of the population (i.e. total income \mytt{PINCP}), but no privacy loss for an attribute that occurs in 50\% of the population (i.e. \mytt{SEX}).

In order to correctly account for the baseline, we measure \textit{Precision Improvement} (PI). This is the improvement in precision above the statistical baseline~\cite{francis2024towards}. PI is measured as:

\begin{equation}
    \text{PI} \ = \ (P_{atk} - P_{base}) / (1 - P_{base})
    \label{eq:pi}
\end{equation}

where $P_{atk}$ and $P_{base}$ are the precision measures for the attack and baseline respectively.

To compute $P_{base}$, we exploit the premise, taken from differential privacy, that a data release that does not include a given individual cannot be regarded as having compromised the privacy of that individual~\cite{Dwork06}. Given this, we can compute $P_{base}$ by splitting the original dataset into \textit{training} and \textit{test} datasets, training an ML model using the quasi-identifier columns as input features, and predicting the value of each target feature in the test dataset. Because the test dataset is not part of the training dataset, any predictions made on the test dataset from the training dataset does not compromise the privacy of individuals in the test dataset. The resulting precision is therefore a privacy-neutral baseline $P_{base}$.

We computed $P_{base}$ for each target column using \mytt{sklearn LogisticRegression} with \mytt{penalty='l1'}, \mytt{C=0.01}, \mytt{solver='saga'}, and \mytt{max\_iter=100}.
\iflong
The results are given in Table~\ref{tab:baselines}.
\else
\fi
The mean $P_{base}$ is 0.61 with a standard deviation of 0.37. The max $P_{base}$ has perfect precision. This is for the column \mytt{DENSITY}, which gives the population density of a \mytt{PUMA} area, is non-personal, and is entirely predicted by \mytt{PUMA}. Given these relatively high baseline precisions, the suggestion of SDNIST that 50\% average precision is safe turns out to be too conservative.

\iflong

            \begin{table}
                \centering
                \begin{tabular}{rl@{\hskip 3pt}|@{\hskip 3pt}rl@{\hskip 3pt}|@{\hskip 3pt}rl@{\hskip 3pt}|@{\hskip 3pt}rl}
                \toprule
        
                PWGTP & 0.01 & AGEP & 0.07 &  WGTP & 0.08 & PINCP & 0.26 \\ 
            
                PINCP\_DECILE & 0.40 & POVPIP & 0.42 &  NPF & 0.52 & INDP & 0.59 \\ 
            
                NOC & 0.69 & DPHY & 0.90 &  DREM & 0.93 & DEAR & 0.97 \\ 
            
                DEYE & 0.97 & HOUSING\_TYPE & 0.98 &  DVET & 0.99 & DENSITY & 1.00 \\ 
            
            \bottomrule
            \end{tabular}
               \caption{Baseline Inference Precision}
            \label{tab:baselines}
            \end{table}
        
\else
\fi

These results are shown in Table~\ref{tab:privacy} under \textit{QI Match}. Figure~\ref{fig:attackPrecCovPairs} plots the median and max PI for each technique. Coverage is the fraction of rows that are unique quasi-identifier matches. $\text{PI} > 0$ leaks some amount of privacy. We believe that $\text{PI} < 0.5$ can conservatively be regarded as anonymous (even at full Coverage). $\text{PI} < 0.5$ gives an attacker substantial uncertainty, and gives individuals substantial deniability, relative to the baseline.

From Figure~\ref{fig:attackPrecCovPairs}, we see that most of the techniques fall below $\text{PI} = 0.5$, and so can be regarded as anonymous. Four techniques are well above $\text{PI} = 0.5$, and so we believe that these can be regarded as not adequately anonymous. These are labeled as `weak' in Table~\ref{tab:infotable}.

\begin{figure}
\begin{center}
\includegraphics[width=1.0\linewidth]{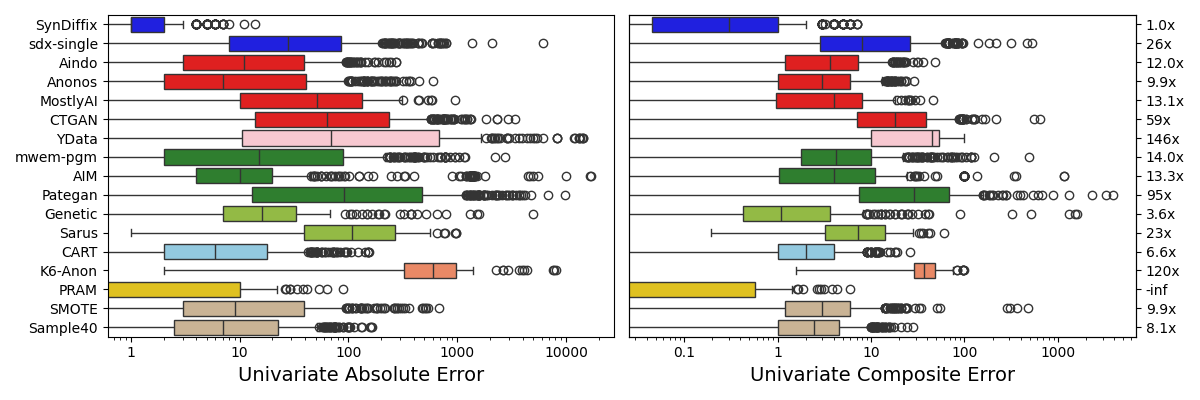}
\caption{Absolute and composite error for univariate (single feature) counts. The composite error is the minimum of the absolute error and the percent relative error. The values give the median composite error Improvement Factor (IF) for SynDiffix. Box plots show 0, 25, 50, 75, and 100 percentiles plus outliers. Note log scale.}
\label{fig:uniStats}
\end{center}
\end{figure}

In addition to the above inference precision measure, SDNIST counts the number of unique full records that are common to both the original and synthetic data. These results are shown in Table~\ref{tab:privacy} as \textit{Full Match}. This does not represent a very meaningful measure of privacy, because there is nothing special about all columns being unique matches versus some fraction of columns (i.e. quasi-identifiers as above or any other set of columns) being unique matches. One can find thousands of unique matches among subsets of columns. What is important is whether inferences well above a precision baseline can be made. Nevertheless, SDNIST tabulates the information, and so we repeat it for completeness.

\subsection{Univariate accuracy}
\label{sec:univariate}

SDNIST measures the count of each feature value (the number of records in which each feature value appears). We define univariate absolute error for each feature value as $E_{abs}= \lvert C_o-C_s \rvert$, where $C_o$ is the count in the original data, and $C_s$ is the count in the synthetic data. We also define \textit{composite error} as $E_{comp}=min(E_{abs},E_{rel})$, where $E_{rel}=100* \lvert C_o-C_s \rvert/C_o$, the percent relative error. Composite error reflects the fact that even large absolute errors can be small relative errors, and vice versa.

Figure~\ref{fig:uniStats} shows the univariate results, with selected corresponding values in Table~\ref{tab:accuracy}. Compared to all other algorithms except PRAM, SynDiffix is multiple times more accurate. (PRAM unfortunately has unacceptable privacy, see\S\ref{sec:privacy}.) The values on the right edge of the composite error plot gives the IF for the median composite error.

SynDiffix is a full order of magnitude more accurate than the best comparable model, the proprietary product Ananos.

\iflong
Looking at absolute error, we see that SynDiffix' univariate counts are almost always within 2 of the true count, and never more than 11 within the true count. Looking at the composite error, we see that 75\% of SynDiffix counts have less that 1\% error. And almost all counts are within 3\% (or have an absolute error of 3 or less).

To get a sense of what an order of magnitude univariate accuracy improvement means, Figure~\ref{fig:uni_pair_example} compares SynDiffix with the best-scoring comparable technique Ananos. While the SynDiffix bars are almost indistinguishable from the true counts, the error in Ananos' counts is clearly visible.

\begin{figure}
\begin{center}
\includegraphics[width=1.0\linewidth]{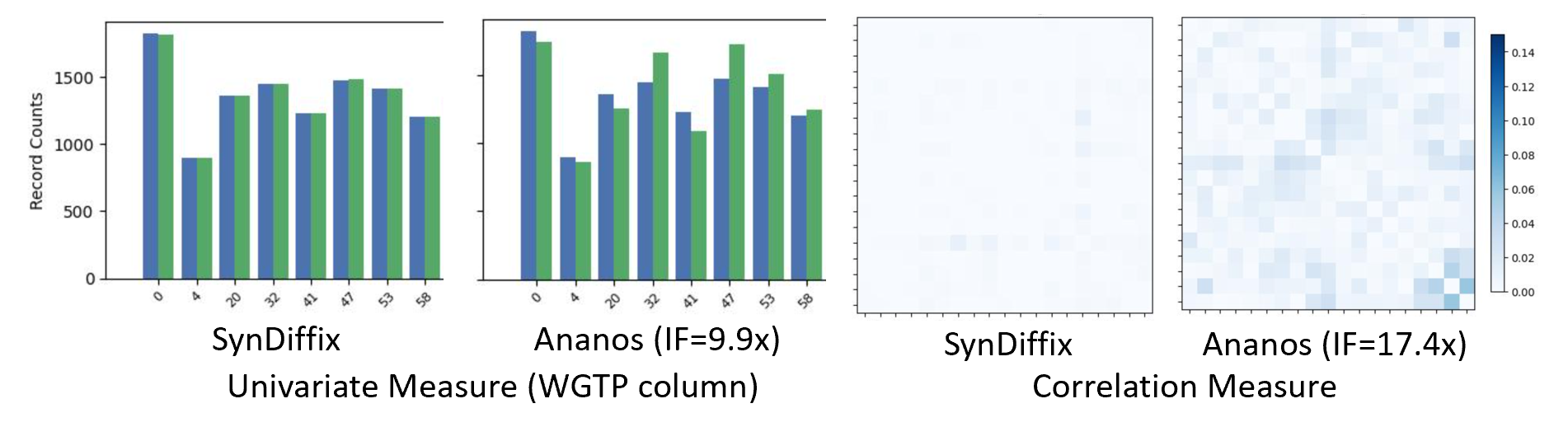}
\caption{Comparison of SynDiffix and Ananos for univariate counts and pairwise correlation. These figures are taken verbatim from the SDNIST summary reports.}
\label{fig:uni_pair_example}
\end{center}
\end{figure}

\else

\fi

\subsection{Pairs accuracy (correlations)}
\label{sec:pairs}

\begin{figure}
\begin{center}
\includegraphics[width=1.0\linewidth]{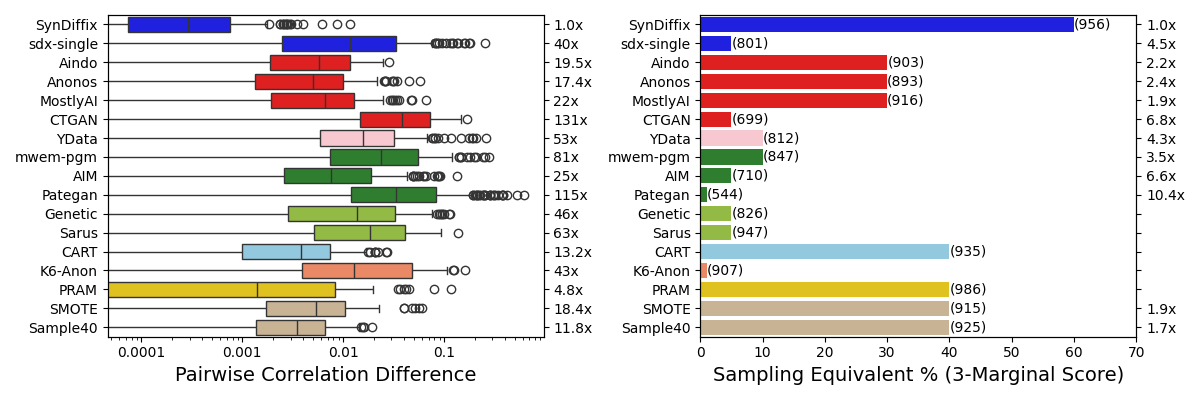}
\caption{Accuracy of pairwise correlations and 3-marginals. The left plot gives the difference between the original and synthetic data for the Kendall Tau correlation coefficient, and corresponding improvement factors (right y-axis). The right plot gives the sampling rate over the original data that would be required to match the 3-marginal accuracy of the synthetic data. The right y-axis is the improvement factor of the 3-marginal accuracy.}
\label{fig:pairTripleStats}
\end{center}
\end{figure}

SDNIST measures pairwise correlations and computes the difference between the original and synthetic data. The left plot of Figure~\ref{fig:pairTripleStats} displays this difference for every feature pair for the Kendall Tau correlation measure. It also shows the median IF. The corresponding numbers are given in Table~\ref{tab:accuracy}. Here we see that SynDiffix is 17x more accurate than the best comparable technique, the proprietary product Ananos.

\iflong

Figure~\ref{fig:uni_pair_example} compares the correlation heatmaps of SynDiffix and Ananos. The quality difference is clearly visible.

\else
\fi

\subsection{3-marginal accuracy}
\label{sec:triples}

SDNIST measures the accuracy of a random selection of 3-feature combinations (3-marginals). SDNIST measures density difference of 232 3-marginals (out of a possible 2024, given 24 features), and derives an average across all measures which it calls the k-marginal score. The score ranges from 0 (no match) to 1000 (perfect match). SDNIST doesn't report the individual measures.

SDNIST also measures k-marginal scores for the original data given different sampling rates. By comparing the k-marginals for synthetic data and sampled original data, SDNIST establishes an equivalence data quality between synthesis and sampling.

The right-hand plot of Figure~\ref{fig:pairTripleStats} shows the sampling equivalent and corresponding k-marginals score (number in parentheses). By this measure, SynDiffix is equivalent to a 60\% sampling rate. The best generative models have a 30\% equivalent sampling rate, and the best differential privacy techniques have a 10\% equivalent sampling rate. Note that the k-marginal scores are not directly comparable between synthetic datasets with a different number of features. In particular, the k-marginal score for PRAM with 10 features (986) exceeds that of SynDiffix with 24 features (956) even though the equivalent sampling rate of PRAM is lower.

\subsection{Linear Regression (four features)}
\label{sec:linear}

SDNIST measures the quality of a linear regression looking at the relationship between educational attainment and relative salary for different Race/Sex groups. An important attribute of a linear regression is the slope, which reflects the relationship between the two variables. We are therefore interested in the amount of error between the slope computed for the original data, and that of the synthetic data. We measure error for a single regression as $E_{reg}= \lvert S_o-S_s \rvert $, where $S_o$ is the slope for the original data, and $S_s$ is the slope for the synthetic data.

\begin{figure}
\begin{center}
\includegraphics[width=1.0\linewidth]{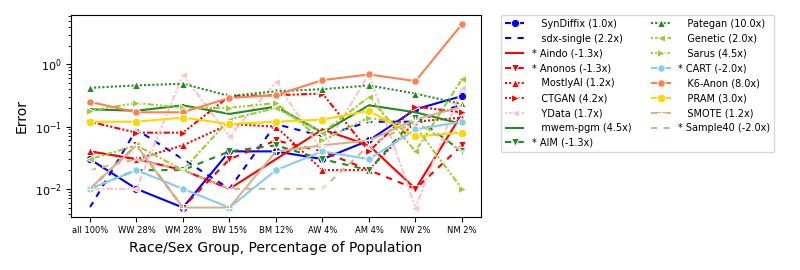}
\caption{Accuracy of the slope of a linear regression on educational attainment and salary. The two-letter code is Race/Sex, where Race is \textbf{W}hite, \textbf{B}lack, \textbf{A}sian, or \textbf{N}ative, and Sex is \textbf{M} or \textbf{W}. The legend gives the improvement factor of SynDiffix over other techniques. Items marked with '*' have less error than SynDiffix. Note the y axis is log scale.}
\label{fig:regression}
\end{center}
\end{figure}

Figures~\ref{fig:regression} plots the error $E_{reg}$ for each of the Race/Sex groups for each technique. It also gives the improvement factor of the median error for SynDiffix over the other techniques, where improvement factor is computed as in Section~\ref{sec:univariate}. 

As Figure~\ref{fig:regression} shows, three comparable techniques have better median accuracy than SynDiffix, by 30\%. CART, which is not comparable because it synthesizes 21 rather than 24 columns, has better accuracy by 2x.

While most of SynDiffix' regressions have less than 6\% error, the two ``Native`` regressions have substantially more error, roughly 10\% and 20\%. The fact that they perform less well is not surprising: the population is smaller (Native/Men NM has 395 out of total 23k records), and more fragmented (``Native'' is actually three separate categories, Hawaiian, Alaskan, and Indian). These two regressions, however, also perform poorly relative to other techniques. As future work, we wish to explore why this might be.

\iflong
To get a sense of the effect of the slope error, Figure~\ref{fig:slopes} shows the original and synthetic regressions curves for SynDiffix and the best comparable technique Ananos. We see that for the larger sample, both regressions are an almost perfect fit. For the small sample, however, Ananos is substantially better.

\begin{figure}
\begin{center}
\includegraphics[width=1.0\linewidth]{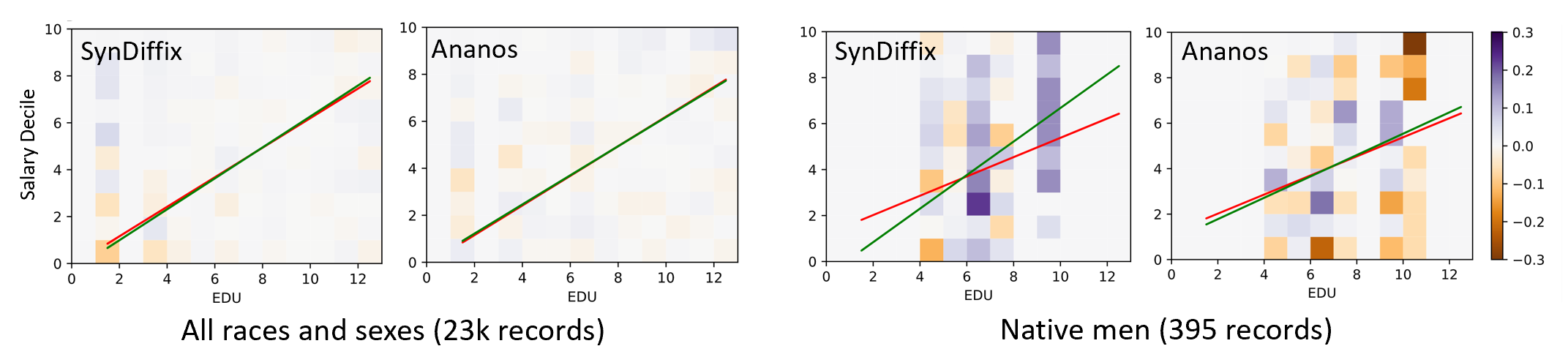}
\caption{Comparison of original (red) and synthetic (green) regressions for SynDiffix and Ananos. The heatmap shows the density difference between the original and synthetic data.}
\label{fig:slopes}
\end{center}
\end{figure}

\else

\fi
\subsection{Propensity mean square error}
\label{sec:pmse}

The Propensity Mean Square Error~\cite{snoke2018general} (PMSE) measures data quality by training a classifier to distinguish between the original and synthetic data. The worse the classifier performance, the higher the synthetic data quality. A PMSE of 0 occurs when the synthetic data perfectly matches the original data.

Snoke et al. proposes that with multi-table synthetic data, a separate PMSE is computed for every table, and the average is taken as the PMSE~\cite{snoke2018general}. We follow that approach here.

\begin{figure}
\begin{center}
\includegraphics[width=1.0\linewidth]{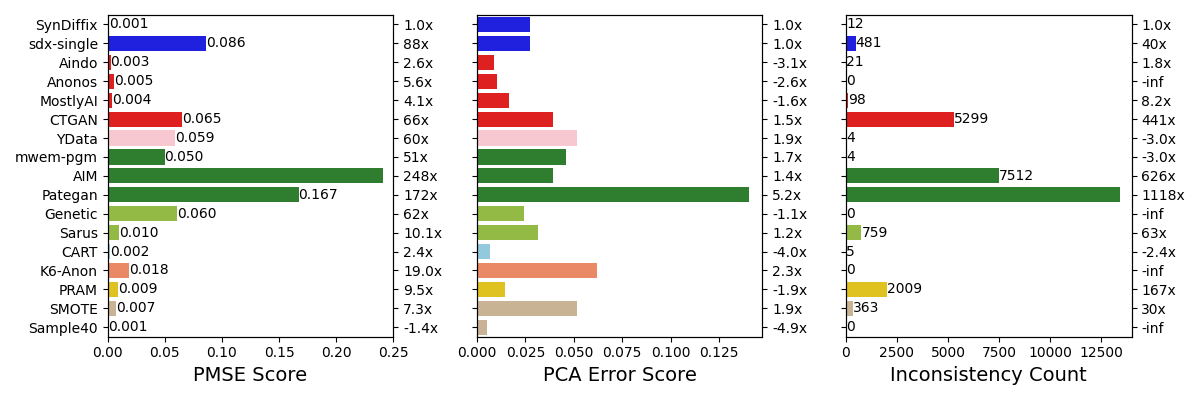}
\caption{Propensity MSE score, PCA error score, and Inconsistency Count. The right y-axes show improvement factor. Details in Table~\ref{tab:pmse}.}
\label{fig:pmsePcaIncon}
\end{center}
\end{figure}

Figure~\ref{fig:pmsePcaIncon} gives the results (with details in Table~\ref{tab:pmse}). SynDiffix has the lowest PMSE score of all techniques except for pure sampling (Sample40). Note that the PMSE score of sdx-single, which measures only the 24-column table, is worse than all of the generative models.

\subsection{Principle Component Analysis (all features)}
\label{sec:pca}

As a means of measuring synthetic data quality across all features, SDNIST runs a Principal Component Analysis (PCA) on the full dataset. SDNIST generates 5 PCs of 5 features each, and then generates a scatterplot visualization of each of 20 PC pairs.  SDNIST zooms in on one particular PC pair (PC-0 and PC-1), highlighting individuals that satisfy a given constraint (marriage status of None versus not None). Note that this is the only SDNIST measure that uses the full 24-column SynDiffix table.

Unfortunately, SDNIST releases only images of the PCs, not the data itself or any accuracy measures. We therefore added a feature to our modified SDNIST implementation to compare the accuracy of the synthetic PCs with the original. Specifically, we compare each of the five original and synthetic PCs using the Kolmogorov-Smirnov score from \mytt{scipy.stats ks\_2samp}, and present the average of these scores in Figure~\ref{fig:pmsePcaIncon} and Table~\ref{tab:pmse}.

We also show the original and synthetic scatterplots side-by-side for in Figure~\ref{fig:pca_grid}. Both visually and from the scores we see that the comparable commercial generative techniques are more accurate than SynDiffix, with Aindo being 3x more accurate than SynDiffix.  In particular, the best scatterplots effectively capture four distinct clusters, where SynDiffix merges two of the clusters together. SynDiffix is more accurate than the DP techniques as well as the open-source generative tool CTGAN.

\subsection{Inconsistencies}
\label{sec:inconsistencies}

SDNIST detects 30 distinct data inconsistencies: combinations of two data values that are impossible. An example of an inconsistency is that a child (age less than 15) can't be a disabled military veteran (columns \mytt{AGEP} and \mytt{DVET}). Each inconsistency is based on two columns, and we use the 2-column tables for this measure.

The results are shown in Figure~\ref{fig:pmsePcaIncon}. SynDiffix has only 12 inconsistencies. Of the comparable techniques, only Ananos and mwem-pgm have fewer (0 and 4 respectively).

\section{Discussion and Conclusion}
\label{sec:conclude}

This paper compares SynDiffix, a new open-source tool for generating structured synthetic data, with 15 other techniques (both commercial and open source), for both privacy and utility metrics. SynDiffix is different from other approaches in that it can safely generate multiple tables, each focused on a given analytic goal. This makes SynDiffix suitable for census data, which commonly releases data as multiple tables.

Using the SDNIST measurement tool modified by us to work with multi-table data, we show that SynDiffix is many times more accurate than other techniques for low-dimension measures, but somewhat worse for high-dimension measures. This suggests that one approach statistics offices may take is to use SynDiffix to release multiple accurate low-dimensional tables, and use a generative technique to release a single full table suitable for ML applications.

While the present study allows for a direct comparison between techniques, it does not evaluate the efficacy of SynDiffix for real analytic use cases. Doing so is an important step in determining whether SynDiffix can serve as a replacement for existing statistical disclosure methods. We hope that the present study motivates work in this direction.

\bibliographystyle{abbrv}
\bibliography{../../masterBib/master}

\newpage
\appendix

\section{Additional data}

This appendix supplies additional data for the various measures. The software that produced these is at \href{https://github.com/yoid2000/sdnist-summary}{https://github.com/yoid2000/sdnist-summary}.

For Table~\ref{tab:privacy}, \textit{count} is the number of unique matches between the original and synthetic data (for full matches and quasi-identifying matches respectively). The percent of full matches relative to the total number of rows is show as \%. \textit{PI} is the inference precision improvement over the statistical baseline. Both median and max PI are shown. The coverage (\textit{cov}) is the median fraction of unique quasi-identifying matches relative to all rows. The QI Match count for SynDiffix is the median count.

For Table~\ref{tab:accuracy}, \textit{N} is the number of datapoints used for the corresponding boxplots, and \textit{med} is the median error (composite in the case of Univariate). \textit{IF} is the improvement factor of the median error. The equivalent sampled original table percentage is \textit{samp}.

For Table~\ref{tab:pmse}, \textit{pmse} and \textit{err} are the PMSE and PCA errors respectively. \textit{count} is the number of inconsistencies. \textit{IF} are the corresponding improvement factors.

\begin{table}
    \centering
    \caption{Summary table for privacy measures.}
    \label{tab:privacy}
    \small
    \begin{tabular}{llrr@{\hskip 14pt}r@{\hskip 6pt}l@{\hskip 6pt}rr}
        \toprule
          &   & \multicolumn{2}{c}{Full Match (\S\ref{sec:privacy})} & \multicolumn{4}{c}{QI Match (\S\ref{sec:privacy})} \\
        
 & & count & \quad \% & med PI & max PI & cov & count \\
\midrule
        \cellcolor{blue} & \href{https://htmlpreview.github.io/?https://github.com/yoid2000/sdnist-summary/blob/main/results/syndiffix_all/report.html}{SynDiffix} & 4 & \quad0.01 & -0.31 & 0.11 & 0.031 & 840.0 \\
        \cellcolor{blue} & \href{https://htmlpreview.github.io/?https://github.com/yoid2000/sdnist-summary/blob/main/results/sdx-single/report.html}{sdx-single} & 4 & \quad0.01 & -0.34 & 0.03 & 0.042 & 1131.0 \\
        \cellcolor{red} & \href{https://htmlpreview.github.io/?https://github.com/yoid2000/sdnist-summary/blob/main/results/aindo_synth/report.html}{Aindo} & 10 & \quad0.04 & -0.34 & 0.21 & 0.060 & 1625.0 \\
        \cellcolor{red} & \href{https://htmlpreview.github.io/?https://github.com/yoid2000/sdnist-summary/blob/main/results/anonos_sdk/report.html}{Anonos} & 5 & \quad0.02 & -0.27 & 0.44 & 0.060 & 1635.0 \\
        \cellcolor{red} & \href{https://htmlpreview.github.io/?https://github.com/yoid2000/sdnist-summary/blob/main/results/mostlyai_sd_platform/report.html}{MostlyAI} & 5 & \quad0.02 & -0.30 & 0.40 & 0.059 & 1606.0 \\
        \cellcolor{red} & \href{https://htmlpreview.github.io/?https://github.com/yoid2000/sdnist-summary/blob/main/results/sdv_ctgan_epochs1000/report.html}{CTGAN} & 0 & \quad0.00 & -0.59 & 0.47 & 0.030 & 811.0 \\
        \cellcolor{pink} & \href{https://htmlpreview.github.io/?https://github.com/yoid2000/sdnist-summary/blob/main/results/ydata_fabric_synthesizers/report.html}{YData} & 453 & \quad1.66 & -1.17 & 0.82 & 0.039 & 1053.0 \\
        \cellcolor{ForestGreen} & \href{https://htmlpreview.github.io/?https://github.com/yoid2000/sdnist-summary/blob/main/results/mwem_pgm/report.html}{mwem-pgm} & 0 & \quad0.00 & -0.26 & 0.58 & 0.042 & 1135.0 \\
        \cellcolor{ForestGreen} & \href{https://htmlpreview.github.io/?https://github.com/yoid2000/sdnist-summary/blob/main/results/aim_e_10_all/report.html}{AIM} & 0 & \quad0.00 & -0.38 & 0.00 & 0.058 & 1589.0 \\
        \cellcolor{ForestGreen} & \href{https://htmlpreview.github.io/?https://github.com/yoid2000/sdnist-summary/blob/main/results/pategan_n_iter_50_e_10_all/report.html}{Pategan} & 0 & \quad0.00 & -2.15 & 0.00 & 0.029 & 782.0 \\
        \cellcolor{YellowGreen} & \href{https://htmlpreview.github.io/?https://github.com/yoid2000/sdnist-summary/blob/main/results/genetic_sd_e_10_simple/report.html}{Genetic} & 5 & \quad0.02 & 0.00 & 0.00 & 0.000 & 0.0 \\
        \cellcolor{YellowGreen} & \href{https://htmlpreview.github.io/?https://github.com/yoid2000/sdnist-summary/blob/main/results/sarus_sdg_demographic/report.html}{Sarus} & 2087 & \quad7.66 & -2.27 & -0.05 & 0.003 & 77.0 \\
        \cellcolor{SkyBlue} & \href{https://htmlpreview.github.io/?https://github.com/yoid2000/sdnist-summary/blob/main/results/cart_cf21/report.html}{CART} & 654 & \quad2.40 & -0.31 & 0.49 & 0.066 & 1797.0 \\
        \cellcolor{Salmon} & \href{https://htmlpreview.github.io/?https://github.com/yoid2000/sdnist-summary/blob/main/results/k_anonymity_k_6/report.html}{K6-Anon} & 7073 & \quad25.95 & 0.00 & 0.00 & 0.000 & 0.0 \\
        \cellcolor{Goldenrod} & \href{https://htmlpreview.github.io/?https://github.com/yoid2000/sdnist-summary/blob/main/results/pram_default/report.html}{PRAM} & 10160 & \quad37.28 & 0.42 & 1.00 & 0.006 & 169.0 \\
        \cellcolor{Tan} & \href{https://htmlpreview.github.io/?https://github.com/yoid2000/sdnist-summary/blob/main/results/smote_target_marital/report.html}{SMOTE} & 4219 & \quad15.48 & 0.62 & 0.84 & 0.078 & 2131.0 \\
        \cellcolor{Tan} & \href{https://htmlpreview.github.io/?https://github.com/yoid2000/sdnist-summary/blob/main/results/subsample_40pcnt_all/report.html}{Sample40} & 10860 & \quad39.85 & 0.94 & 1.00 & 0.153 & 4169.0 \\
        \bottomrule
    \end{tabular}
\end{table}

\begin{table}
    \centering
    \caption{Summary table for low-dimensional accuracy measures.}
    \label{tab:accuracy}
    \small
    \begin{tabular}{llrlr@{\hskip 10pt}r@{\hskip 6pt}l@{\hskip 6pt}r@{\hskip 10pt}r@{\hskip 6pt}r@{\hskip 6pt}r@{\hskip 6pt}r}
        \toprule
          &   & \multicolumn{3}{c}{Univariate (\S\ref{sec:univariate})} & \multicolumn{3}{c}{Correlation (\S\ref{sec:pairs})} & \multicolumn{4}{c}{3-marginals (\S\ref{sec:triples})} \\
        
 & & N & med & IF & N & med & IF & score & IF & samp & IF \\
\midrule
        \cellcolor{blue} & \href{https://htmlpreview.github.io/?https://github.com/yoid2000/sdnist-summary/blob/main/results/syndiffix_all/report.html}{SynDiffix} & 499 & 0.3 & 1.0x & 190 & 0.0003 & 1.0x & 956 & 1.0x & 60\% & 1.0x \\
        \cellcolor{blue} & \href{https://htmlpreview.github.io/?https://github.com/yoid2000/sdnist-summary/blob/main/results/sdx-single/report.html}{sdx-single} & 510 & 8.0 & 26x & 190 & 0.0117 & 40x & 801 & 4.5x & 5\% & 2.4x \\
        \cellcolor{red} & \href{https://htmlpreview.github.io/?https://github.com/yoid2000/sdnist-summary/blob/main/results/aindo_synth/report.html}{Aindo} & 499 & 3.6 & 12.0x & 190 & 0.0057 & 19.5x & 903 & 2.2x & 30\% & 1.8x \\
        \cellcolor{red} & \href{https://htmlpreview.github.io/?https://github.com/yoid2000/sdnist-summary/blob/main/results/anonos_sdk/report.html}{Anonos} & 499 & 3.0 & 9.9x & 190 & 0.0051 & 17.4x & 893 & 2.4x & 30\% & 1.8x \\
        \cellcolor{red} & \href{https://htmlpreview.github.io/?https://github.com/yoid2000/sdnist-summary/blob/main/results/mostlyai_sd_platform/report.html}{MostlyAI} & 216 & 4.0 & 13.1x & 190 & 0.0066 & 22x & 916 & 1.9x & 30\% & 1.8x \\
        \cellcolor{red} & \href{https://htmlpreview.github.io/?https://github.com/yoid2000/sdnist-summary/blob/main/results/sdv_ctgan_epochs1000/report.html}{CTGAN} & 502 & 18.1 & 59x & 190 & 0.0383 & 131x & 699 & 6.8x & 5\% & 2.4x \\
        \cellcolor{pink} & \href{https://htmlpreview.github.io/?https://github.com/yoid2000/sdnist-summary/blob/main/results/ydata_fabric_synthesizers/report.html}{YData} & 499 & 44.6 & 146x & 153 & 0.0156 & 53x & 812 & 4.3x & 10\% & 2.2x \\
        \cellcolor{ForestGreen} & \href{https://htmlpreview.github.io/?https://github.com/yoid2000/sdnist-summary/blob/main/results/mwem_pgm/report.html}{mwem-pgm} & 499 & 4.3 & 14.0x & 190 & 0.0237 & 81x & 847 & 3.5x & 10\% & 2.2x \\
        \cellcolor{ForestGreen} & \href{https://htmlpreview.github.io/?https://github.com/yoid2000/sdnist-summary/blob/main/results/aim_e_10_all/report.html}{AIM} & 499 & 4.0 & 13.3x & 190 & 0.0075 & 25x & 710 & 6.6x & 5\% & 2.4x \\
        \cellcolor{ForestGreen} & \href{https://htmlpreview.github.io/?https://github.com/yoid2000/sdnist-summary/blob/main/results/pategan_n_iter_50_e_10_all/report.html}{Pategan} & 505 & 29.0 & 95x & 190 & 0.0336 & 115x & 544 & 10.4x & 1\% & 2.5x \\
        \cellcolor{YellowGreen} & \href{https://htmlpreview.github.io/?https://github.com/yoid2000/sdnist-summary/blob/main/results/genetic_sd_e_10_simple/report.html}{Genetic} & 202 & 1.1 & 3.6x & 190 & 0.0136 & 46x & 826 &   & 5\% & 2.4x \\
        \cellcolor{YellowGreen} & \href{https://htmlpreview.github.io/?https://github.com/yoid2000/sdnist-summary/blob/main/results/sarus_sdg_demographic/report.html}{Sarus} & 77 & 7.2 & 23x & 55 & 0.0186 & 63x & 947 &   & 5\% & 2.4x \\
        \cellcolor{SkyBlue} & \href{https://htmlpreview.github.io/?https://github.com/yoid2000/sdnist-summary/blob/main/results/cart_cf21/report.html}{CART} & 451 & 2.0 & 6.6x & 171 & 0.0038 & 13.2x & 935 &   & 40\% & 1.5x \\
        \cellcolor{Salmon} & \href{https://htmlpreview.github.io/?https://github.com/yoid2000/sdnist-summary/blob/main/results/k_anonymity_k_6/report.html}{K6-Anon} & 77 & 36.7 & 120x & 55 & 0.0127 & 43x & 907 &   & 1\% & 2.5x \\
        \cellcolor{Goldenrod} & \href{https://htmlpreview.github.io/?https://github.com/yoid2000/sdnist-summary/blob/main/results/pram_default/report.html}{PRAM} & 77 & 0.0 & -inf & 55 & 0.0014 & 4.8x & 986 &   & 40\% & 1.5x \\
        \cellcolor{Tan} & \href{https://htmlpreview.github.io/?https://github.com/yoid2000/sdnist-summary/blob/main/results/smote_target_marital/report.html}{SMOTE} & 499 & 3.0 & 9.9x & 190 & 0.0054 & 18.4x & 915 & 1.9x & 40\% & 1.5x \\
        \cellcolor{Tan} & \href{https://htmlpreview.github.io/?https://github.com/yoid2000/sdnist-summary/blob/main/results/subsample_40pcnt_all/report.html}{Sample40} & 499 & 2.5 & 8.1x & 190 & 0.0035 & 11.8x & 925 & 1.7x & 40\% & 1.5x \\
        \bottomrule
    \end{tabular}
\end{table}

\begin{table}
    \centering
    \caption{Summary table for Propensity MSE, PCA Error, and Inconsistencies.}
    \label{tab:pmse}
    \small
    \begin{tabular}{llr@{\hskip 6pt}r@{\hskip 14pt}r@{\hskip 6pt}r@{\hskip 14pt}rr}
        \toprule
          &   & \multicolumn{2}{c}{PMSE (\S\ref{sec:pmse})} & \multicolumn{2}{c}{PCA Error (\S\ref{sec:pca})} & \multicolumn{2}{c}{Inconsistencies (\S\ref{sec:inconsistencies})} \\
        
 & & pmse & IF & ks-score & IF & count & IF \\
\midrule
        \cellcolor{blue} & \href{https://htmlpreview.github.io/?https://github.com/yoid2000/sdnist-summary/blob/main/results/syndiffix_all/report.html}{SynDiffix} & 0.0010 & 1.0x & 0.0272 & 1.0x & 12 & 1.0x \\
        \cellcolor{blue} & \href{https://htmlpreview.github.io/?https://github.com/yoid2000/sdnist-summary/blob/main/results/sdx-single/report.html}{sdx-single} & 0.0859 & 88x & 0.0272 & 1.0x & 481 & 40x \\
        \cellcolor{red} & \href{https://htmlpreview.github.io/?https://github.com/yoid2000/sdnist-summary/blob/main/results/aindo_synth/report.html}{Aindo} & 0.0025 & 2.6x & 0.0089 & -3.1x & 21 & 1.8x \\
        \cellcolor{red} & \href{https://htmlpreview.github.io/?https://github.com/yoid2000/sdnist-summary/blob/main/results/anonos_sdk/report.html}{Anonos} & 0.0054 & 5.6x & 0.0106 & -2.6x & 0 & -inf \\
        \cellcolor{red} & \href{https://htmlpreview.github.io/?https://github.com/yoid2000/sdnist-summary/blob/main/results/mostlyai_sd_platform/report.html}{MostlyAI} & 0.0040 & 4.1x & 0.0166 & -1.6x & 98 & 8.2x \\
        \cellcolor{red} & \href{https://htmlpreview.github.io/?https://github.com/yoid2000/sdnist-summary/blob/main/results/sdv_ctgan_epochs1000/report.html}{CTGAN} & 0.0649 & 66x & 0.0394 & 1.5x & 5299 & 441x \\
        \cellcolor{pink} & \href{https://htmlpreview.github.io/?https://github.com/yoid2000/sdnist-summary/blob/main/results/ydata_fabric_synthesizers/report.html}{YData} & 0.0588 & 60x & 0.0518 & 1.9x & 4 & -3.0x \\
        \cellcolor{ForestGreen} & \href{https://htmlpreview.github.io/?https://github.com/yoid2000/sdnist-summary/blob/main/results/mwem_pgm/report.html}{mwem-pgm} & 0.0498 & 51x & 0.0458 & 1.7x & 4 & -3.0x \\
        \cellcolor{ForestGreen} & \href{https://htmlpreview.github.io/?https://github.com/yoid2000/sdnist-summary/blob/main/results/aim_e_10_all/report.html}{AIM} & 0.2412 & 248x & 0.0391 & 1.4x & 7512 & 626x \\
        \cellcolor{ForestGreen} & \href{https://htmlpreview.github.io/?https://github.com/yoid2000/sdnist-summary/blob/main/results/pategan_n_iter_50_e_10_all/report.html}{Pategan} & 0.1670 & 172x & 0.1402 & 5.2x & 13423 & 1118x \\
        \cellcolor{YellowGreen} & \href{https://htmlpreview.github.io/?https://github.com/yoid2000/sdnist-summary/blob/main/results/genetic_sd_e_10_simple/report.html}{Genetic} & 0.0603 & 62x & 0.0243 & -1.1x & 0 & -inf \\
        \cellcolor{YellowGreen} & \href{https://htmlpreview.github.io/?https://github.com/yoid2000/sdnist-summary/blob/main/results/sarus_sdg_demographic/report.html}{Sarus} & 0.0098 & 10.1x & 0.0317 & 1.2x & 759 & 63x \\
        \cellcolor{SkyBlue} & \href{https://htmlpreview.github.io/?https://github.com/yoid2000/sdnist-summary/blob/main/results/cart_cf21/report.html}{CART} & 0.0023 & 2.4x & 0.0068 & -4.0x & 5 & -2.4x \\
        \cellcolor{Salmon} & \href{https://htmlpreview.github.io/?https://github.com/yoid2000/sdnist-summary/blob/main/results/k_anonymity_k_6/report.html}{K6-Anon} & 0.0184 & 19.0x & 0.0619 & 2.3x & 0 & -inf \\
        \cellcolor{Goldenrod} & \href{https://htmlpreview.github.io/?https://github.com/yoid2000/sdnist-summary/blob/main/results/pram_default/report.html}{PRAM} & 0.0093 & 9.5x & 0.0144 & -1.9x & 2009 & 167x \\
        \cellcolor{Tan} & \href{https://htmlpreview.github.io/?https://github.com/yoid2000/sdnist-summary/blob/main/results/smote_target_marital/report.html}{SMOTE} & 0.0070 & 7.3x & 0.0518 & 1.9x & 363 & 30x \\
        \cellcolor{Tan} & \href{https://htmlpreview.github.io/?https://github.com/yoid2000/sdnist-summary/blob/main/results/subsample_40pcnt_all/report.html}{Sample40} & 0.0007 & -1.4x & 0.0055 & -4.9x & 0 & -inf \\
        \bottomrule
    \end{tabular}
\end{table}

\newcolumntype{M}[1]{>{\centering\arraybackslash}m{#1}}
\begin{figure}[p!] % Place the grid on a separate page
\centering
\begin{tabular}{@{} M{0.166\textwidth} M{0.166\textwidth} | M{0.166\textwidth} M{0.166\textwidth} | M{0.166\textwidth} M{0.166\textwidth} @{}}
       \includegraphics[width=\linewidth]{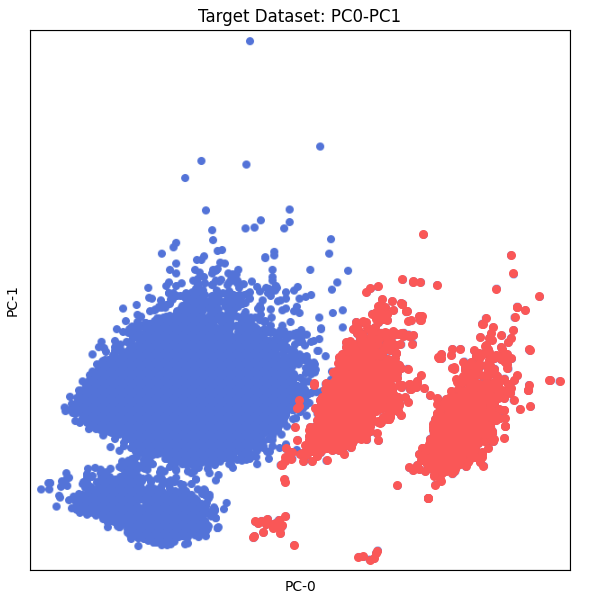} &
       \includegraphics[width=\linewidth]{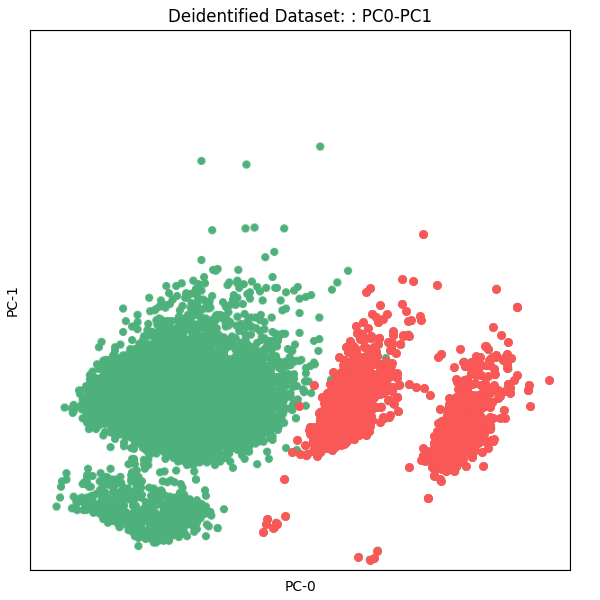} &
       \includegraphics[width=\linewidth]{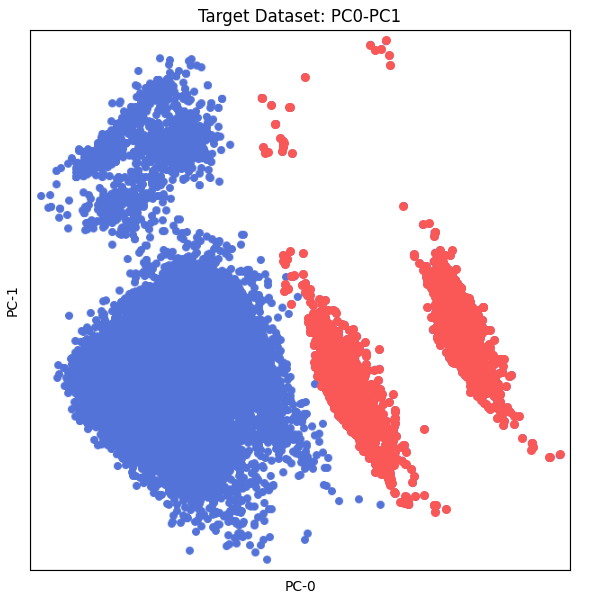} &
       \includegraphics[width=\linewidth]{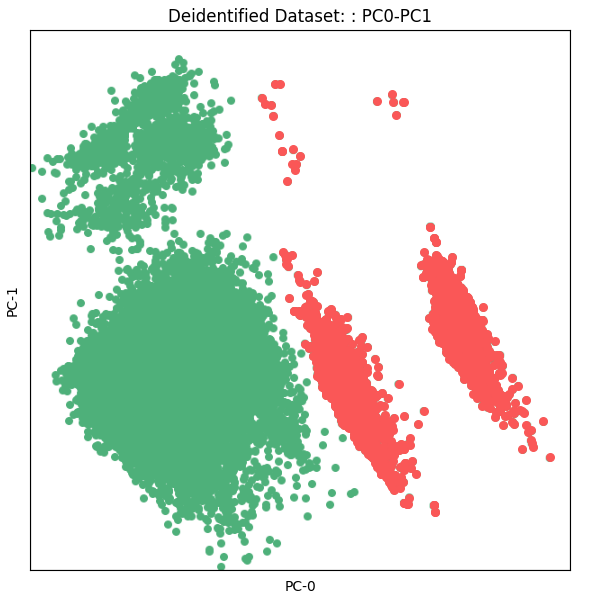} &
       \includegraphics[width=\linewidth]{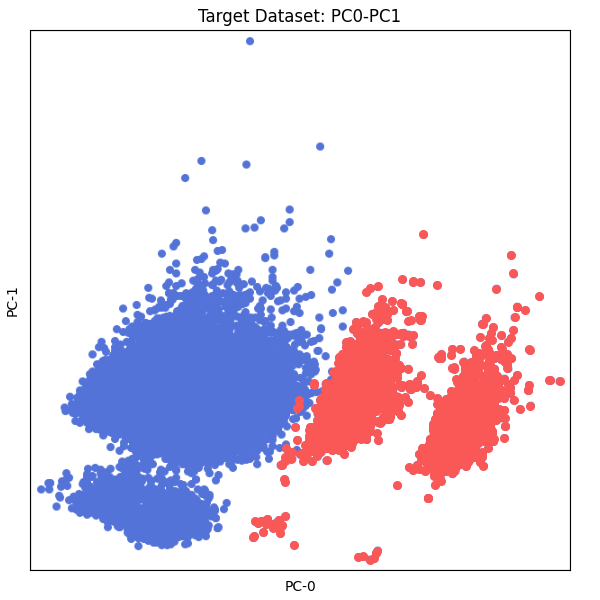} &
       \includegraphics[width=\linewidth]{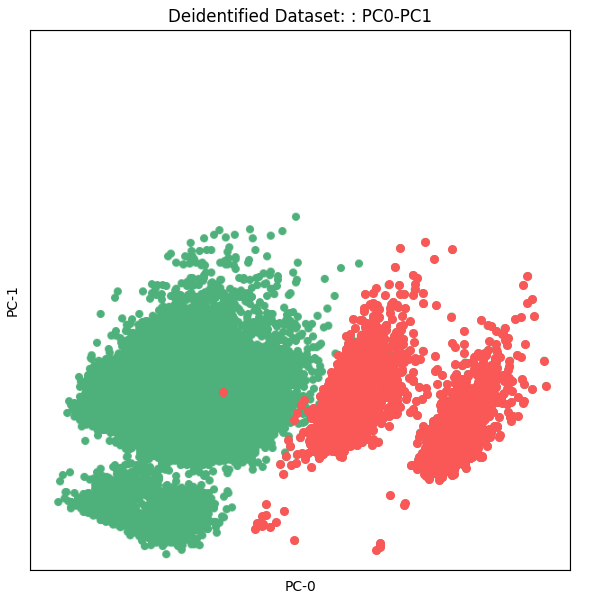} \\ 
\multicolumn{2}{c|}{Sample40, 0.0055} &
\multicolumn{2}{c|}{CART, 0.0068} &
\multicolumn{2}{c}{Aindo, 0.0089} \\ 
 \hline 
       \includegraphics[width=\linewidth]{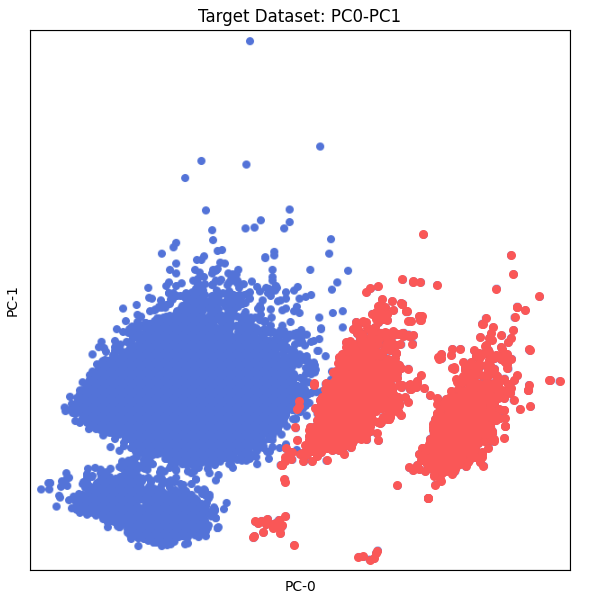} &
       \includegraphics[width=\linewidth]{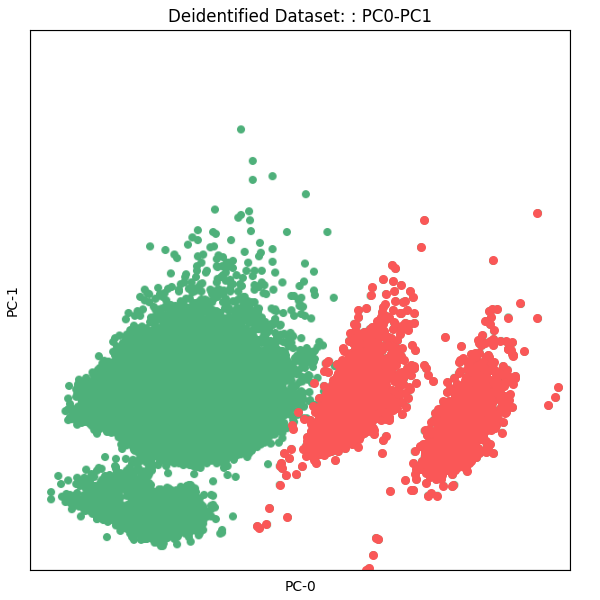} &
       \includegraphics[width=\linewidth]{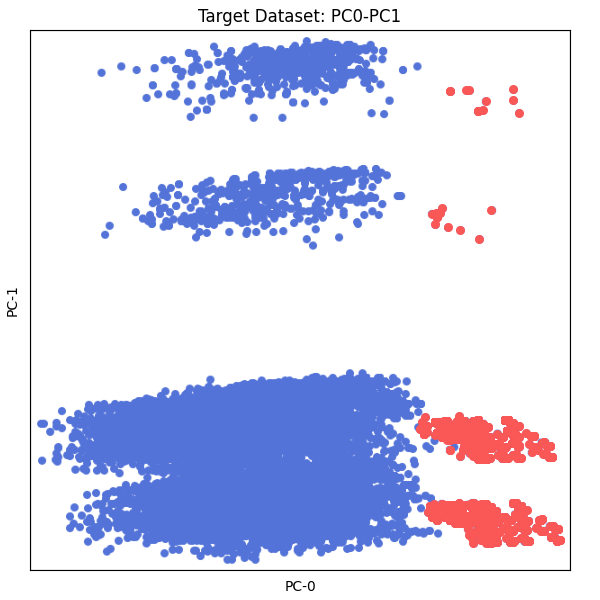} &
       \includegraphics[width=\linewidth]{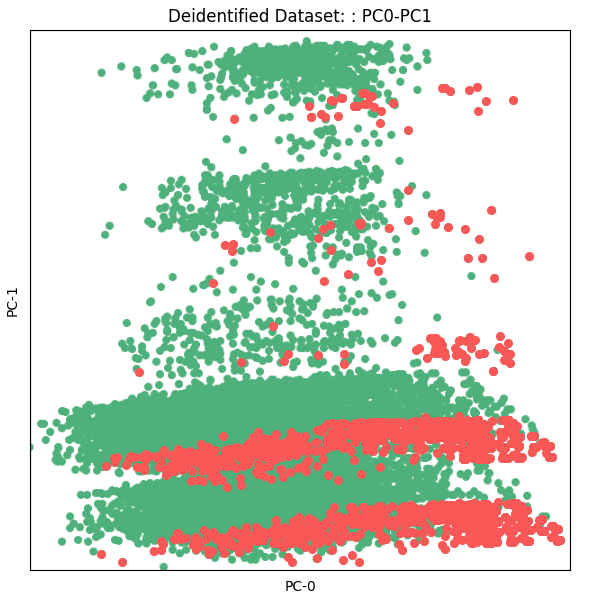} &
       \includegraphics[width=\linewidth]{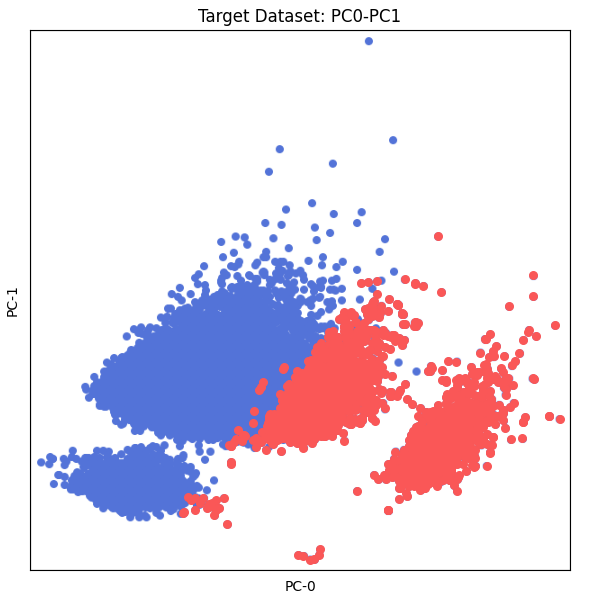} &
       \includegraphics[width=\linewidth]{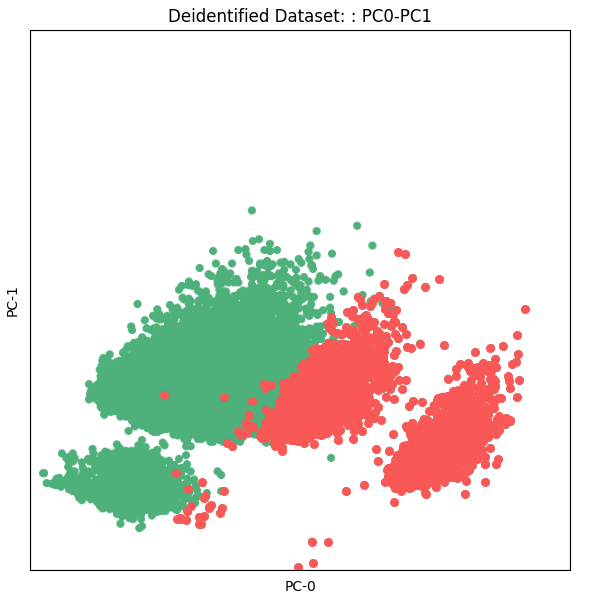} \\ 
\multicolumn{2}{c|}{Anonos, 0.0106} &
\multicolumn{2}{c|}{PRAM, 0.0144} &
\multicolumn{2}{c}{MostlyAI, 0.0166} \\ 
 \hline 
       \includegraphics[width=\linewidth]{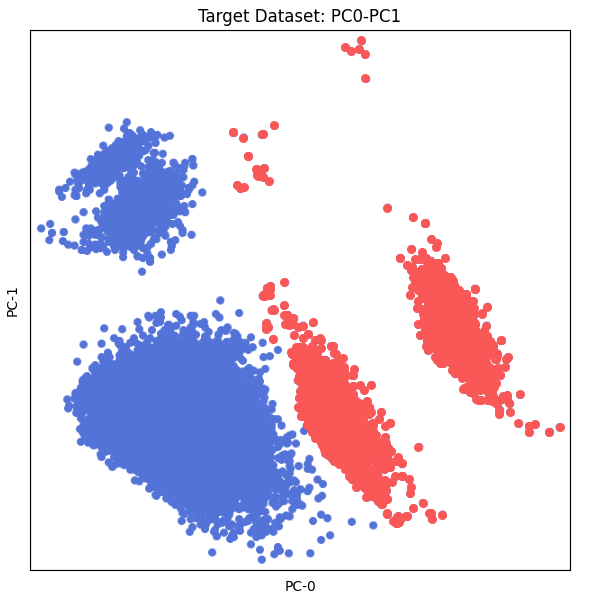} &
       \includegraphics[width=\linewidth]{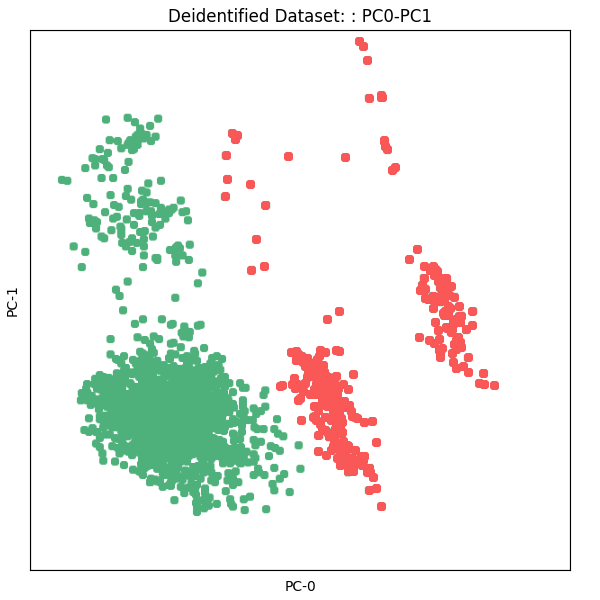} &
       \includegraphics[width=\linewidth]{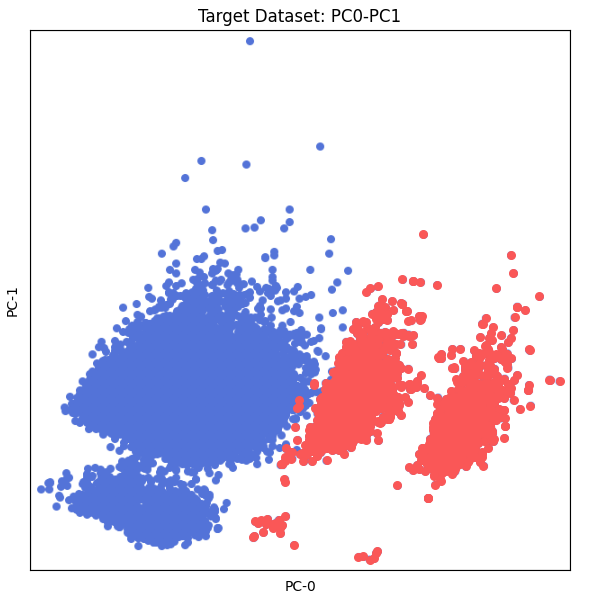} &
       \includegraphics[width=\linewidth]{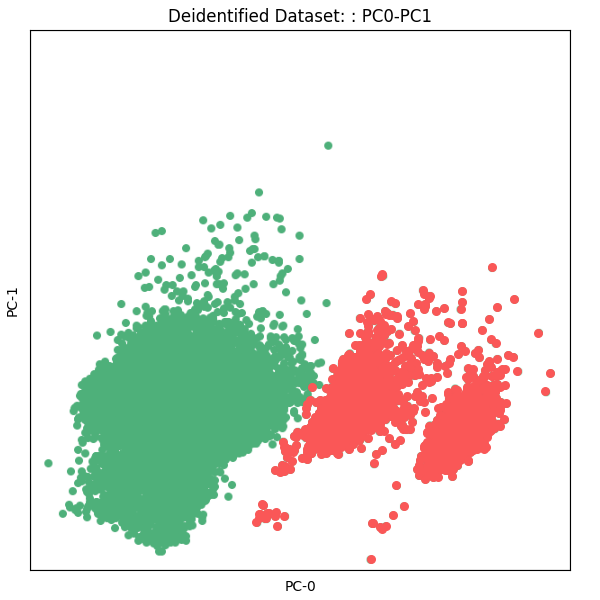} &
       \includegraphics[width=\linewidth]{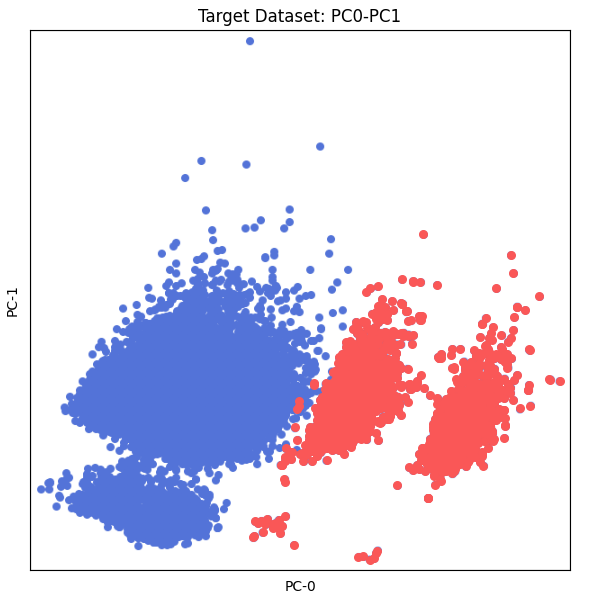} &
       \includegraphics[width=\linewidth]{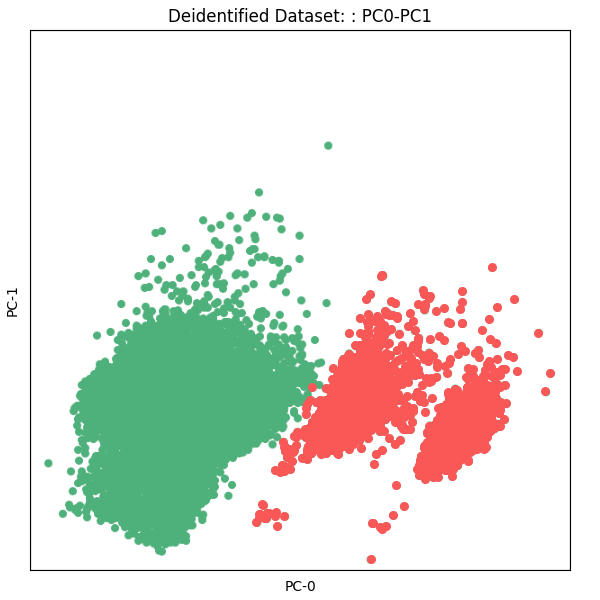} \\ 
\multicolumn{2}{c|}{Genetic, 0.0243} &
\multicolumn{2}{c|}{SynDiffix, 0.0272} &
\multicolumn{2}{c}{sdx-single, 0.0272} \\ 
 \hline 
       \includegraphics[width=\linewidth]{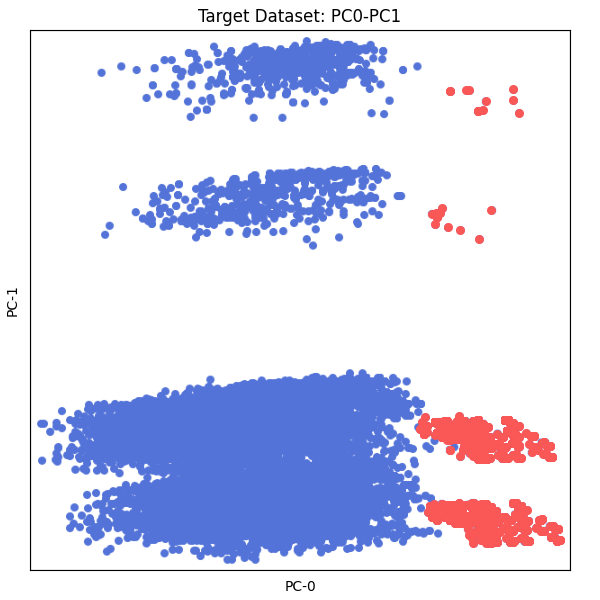} &
       \includegraphics[width=\linewidth]{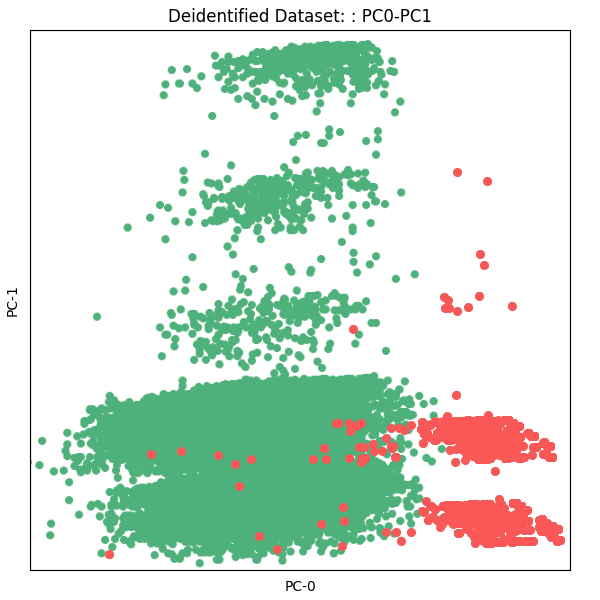} &
       \includegraphics[width=\linewidth]{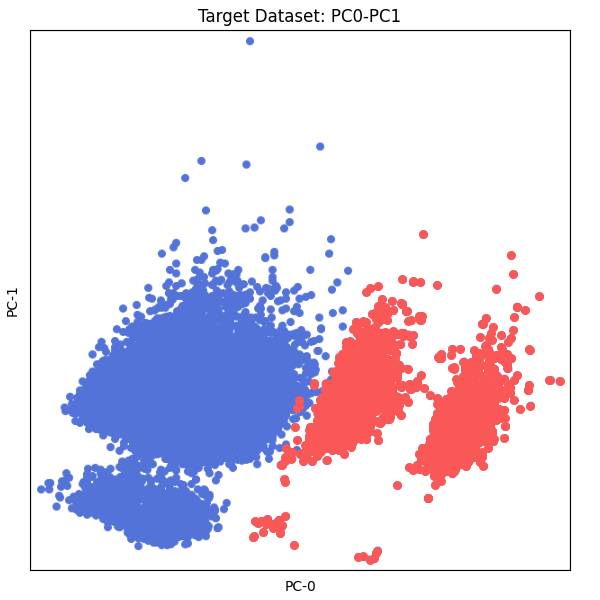} &
       \includegraphics[width=\linewidth]{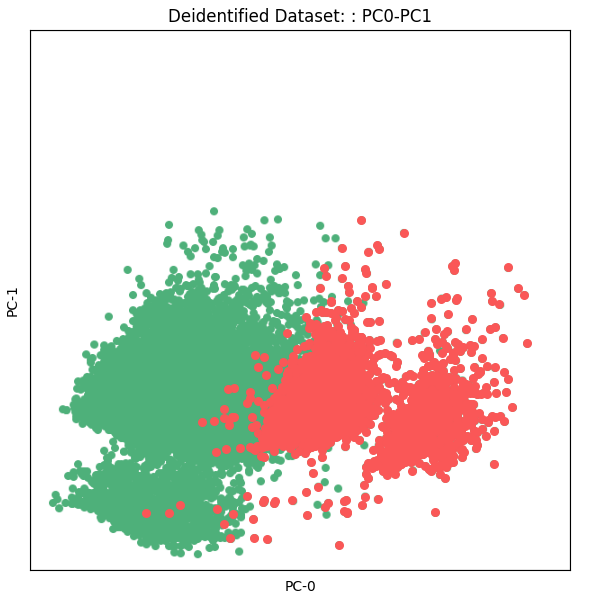} &
       \includegraphics[width=\linewidth]{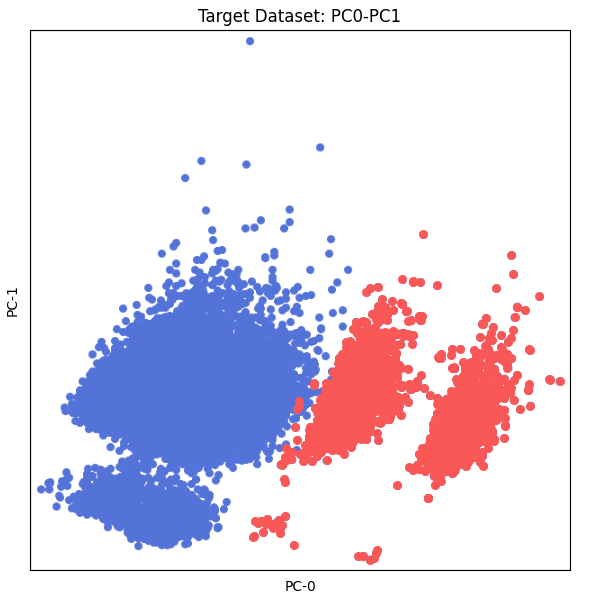} &
       \includegraphics[width=\linewidth]{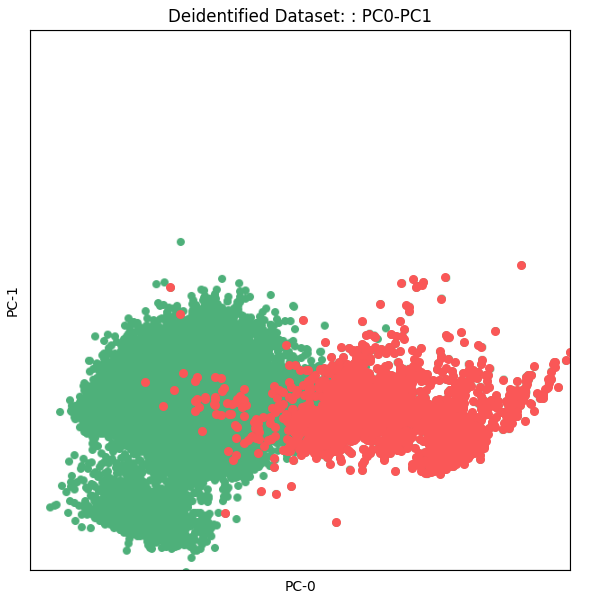} \\ 
\multicolumn{2}{c|}{Sarus, 0.0317} &
\multicolumn{2}{c|}{AIM, 0.0391} &
\multicolumn{2}{c}{CTGAN, 0.0394} \\ 
 \hline 
       \includegraphics[width=\linewidth]{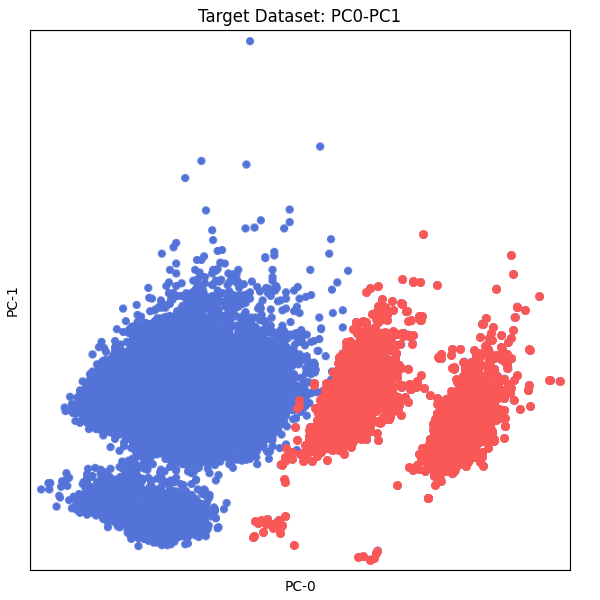} &
       \includegraphics[width=\linewidth]{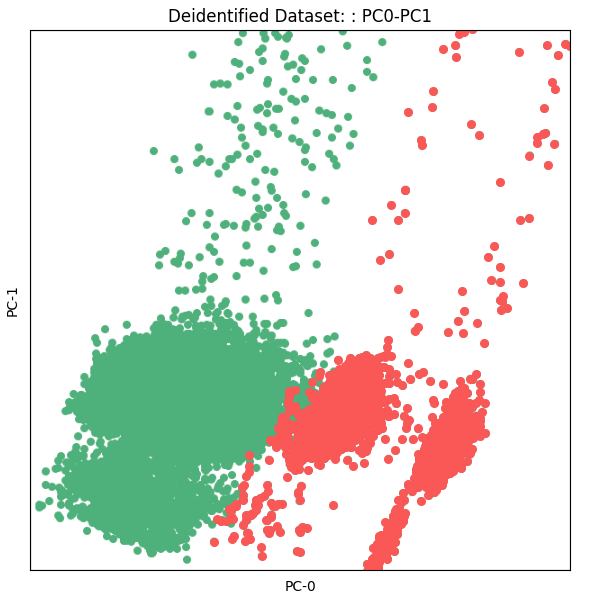} &
       \includegraphics[width=\linewidth]{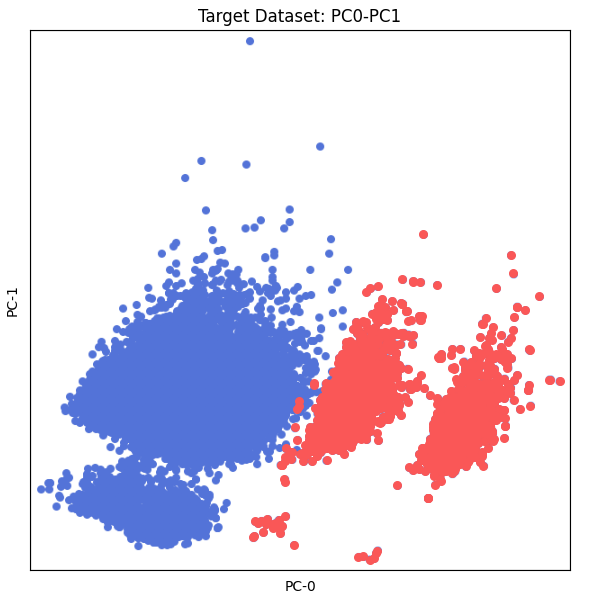} &
       \includegraphics[width=\linewidth]{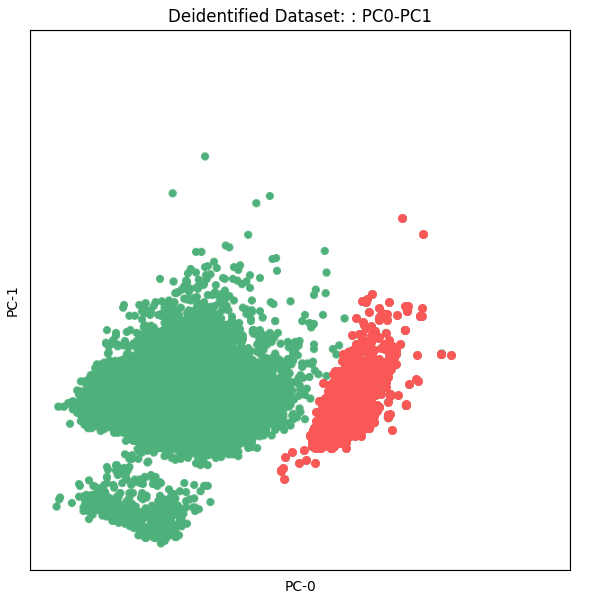} &
       \includegraphics[width=\linewidth]{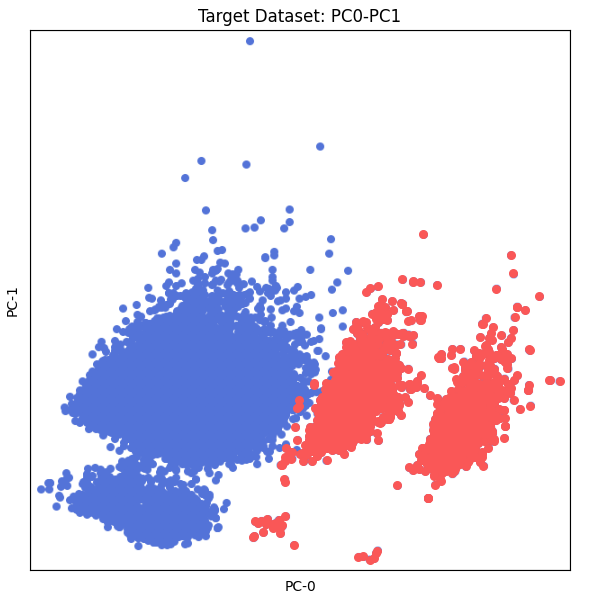} &
       \includegraphics[width=\linewidth]{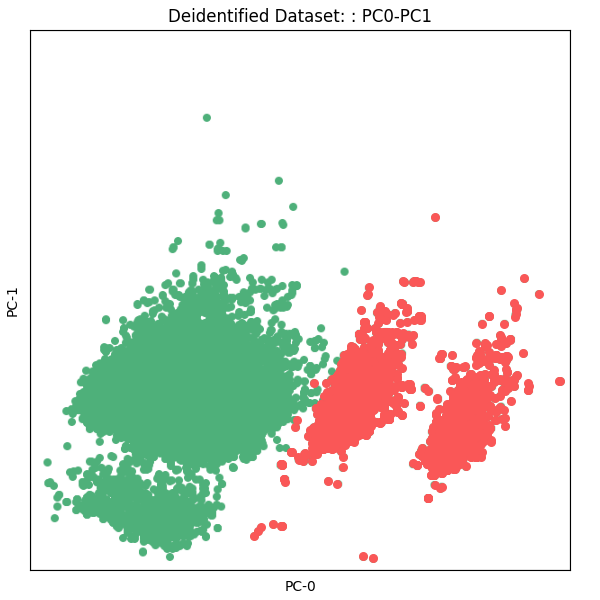} \\ 
\multicolumn{2}{c|}{mwem-pgm, 0.0458} &
\multicolumn{2}{c|}{YData, 0.0518} &
\multicolumn{2}{c}{SMOTE, 0.0518} \\ 
 \hline 
       \includegraphics[width=\linewidth]{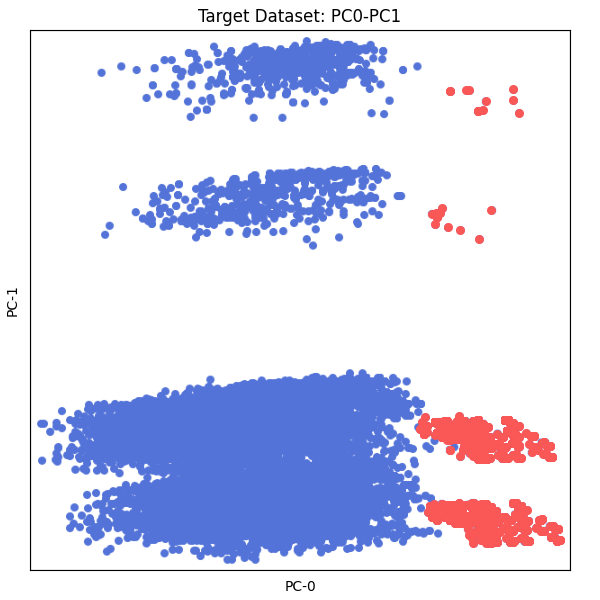} &
       \includegraphics[width=\linewidth]{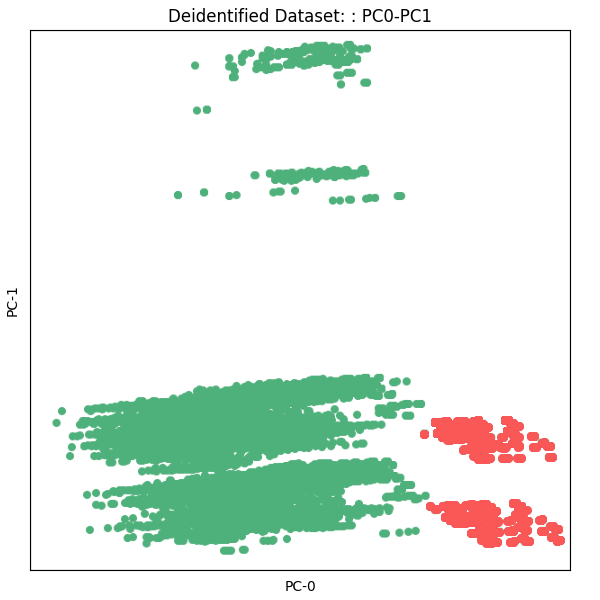} &
       \includegraphics[width=\linewidth]{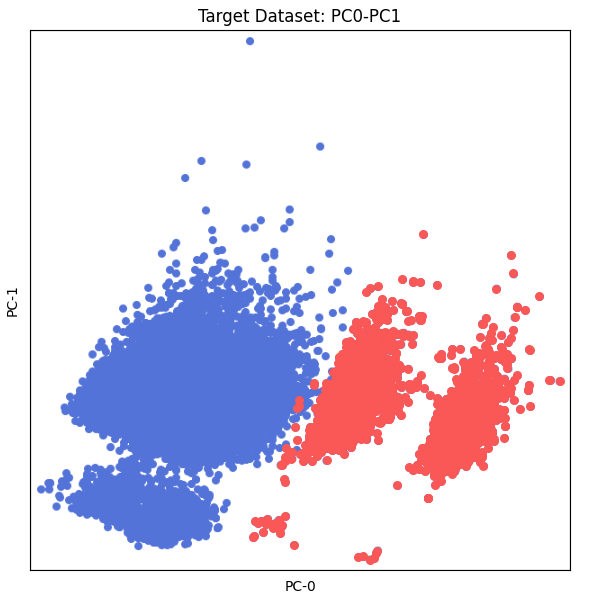} &
       \includegraphics[width=\linewidth]{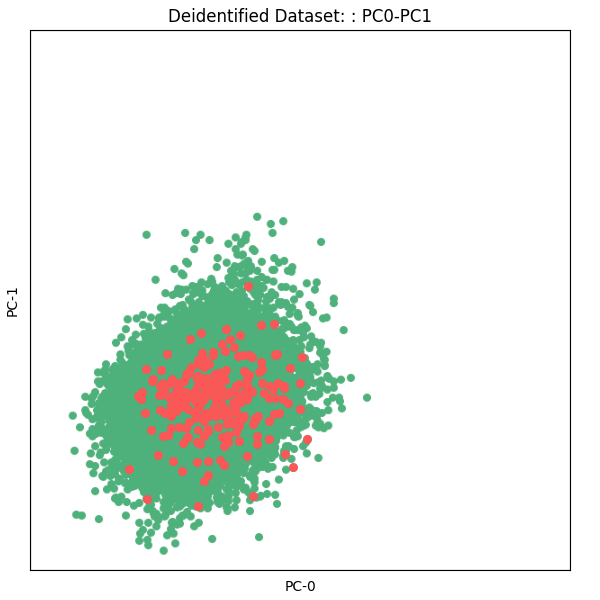} \\ 
\multicolumn{2}{c|}{K6-Anon, 0.0619} &
\multicolumn{2}{c}{Pategan, 0.1402} \\ 
    \end{tabular}
\caption{Original (left) and synthetic (right) scatterplot and average Kolmogorov-Smirnov score for all principle components, ordered by most-to-least accurate.}
\label{fig:pca_grid}
\end{figure}

\end{document}